\documentclass[graybox]{svmult}

\usepackage{type1cm}        
\usepackage{makeidx}         
\usepackage{graphicx}        
\usepackage{multicol}        
\usepackage[bottom]{footmisc}

\makeindex             

\usepackage{lscape}

\newcommand{\afr}{$Af\!\rho$}
\newcommand{\cg}{67P/C-G}
\newcommand{\micron}{$\mu$m}
\newcommand{\Isun} {I_{\odot}}
\newcommand{\df}{{\rm d}}

\newcommand{\degree}{$^{\circ}$}
\newcommand{\rh}{$r_{\rm h}$}

\begin{document}

\title*{Dust Environment Modelling of Comet 67P/Churyumov-Gerasimenko}
\author{Jessica Agarwal,
         Michael M{\"u}ller, and
         Eberhard Gr{\"u}n}
\institute{Jessica Agarwal 
\at ESA Research and Scientific Support Department, ESTEC, Postbus 299, NL-2200 AG Noordwijk,
\email{jagarwal@rssd.esa.int}
\and Michael M{\"u}ller
\at ESA, ESOC, Robert-Bosch-Str. 5, D-64293 Darmstadt,
\email{michael.mueller@esa.int}
\and Eberhard Gr{\"u}n 
\at Max-Planck-Institut f{\"u}r Kernphysik, Saupfercheckweg 1, D-69117
Heidelberg,
\email{eberhard.gruen@mpi-hd.mpg.de}}
%
%
\maketitle

\abstract{Dust is an important constituent in cometary comae; 
its analysis is one of the major objectives of ESA's
Rosetta mission to comet 67P/Churyu\-mov-Gera\-si\-menko (C--G). Several instruments
aboard Rosetta are dedicated to studying various aspects of dust in the
cometary coma, all of which require a certain level of exposure to dust to achieve their goals. At the same time, impacts of dust particles can
constitute a hazard to the spacecraft. To conciliate the demands of
dust collection instruments and spacecraft safety, it is desirable to assess
the dust environment in the coma even before the
arrival of Rosetta. 
We describe the present status of modelling the dust coma of 67P/C--G and 
predict 
the speed and flux of dust in the coma, the dust fluence
on a spacecraft along sample trajectories, and the radiation environment in
the coma. The model will need to be refined when more details of the coma are revealed by observations.
An overview of astronomical observations of 67P/C--G is given and
model parameters are derived from these data where possible.
For quantities not yet measured for 67P/C--G, we use values obtained for other
comets. One of the most
important and most controversial parameters is the dust mass distribution. We
summarise the mass distribution functions derived from the in-situ
measurements at 
comet 1P/Halley in 1986. For 67P/C--G,
constraining the mass distribution is currently only possible by the analysis of
astronomical images. We find that the results from such analyses are at
present rather heterogeneous, and we identify a need to find a model that is
reconcilable with all available observations.}

\section{Introduction}
\label{sec:intro}
The inner comae of comets are the most dust rich environments
in the solar system. Almost everything we see from a comet with the naked eye
is dust. Both the coma and the tail are seen as
sunlight scattered by \micron-sized dust. Meteors, especially in meteor
streams, are caused by mm- to cm-sized particles that originate to a
large extent in comets. Fireballs are due to multi-tons
boulders some of which are believed to stem from comets as
well. 

Spacecraft missions to comets must take careful
precautions to survive the hazards of the cometary
dust cloud. Special dust shields have been designed to protect
the spacecraft. Nevertheless, the dust environment can have
detrimental effects on some aspects of the missions. The
Giotto spacecraft flew by comet 1P/Halley at a distance of 600 km
and a speed of 70 km/s. During this close fly-by Giotto was hit
by mm-sized dust particles. The impacts caused nutation of the
spacecraft spin, and the data transmission was disrupted. Some
experiment sensors suffered damage during this fly-by.
Similarly, the 350-kg impactor of the Deep Impact probe experienced
attitude disturbances from dust grains before it hit the
nucleus of comet 9P/Tempel~1. 
Owing to the low velocity of Rosetta relative to the comet, the consequences
of dust impacts will be much less severe than for the fly-by
missions. But detailed knowledge of the
dust environment will be vital for the planning of spacecraft operations 
at the comet and is thus of crucial importance to optimise Rosetta's
scientific return.

One of the problems in characterising the dust environment of a
comet is that information on the nucleus, its dust and gas
release is very limited. 
Before 1986, observations of cometary
dust was the domain of astronomers. The method of Finson and Probstein \cite{finson-probstein1968a}
was the first to be used to determine the size distribution
of dust from observations of the tail at
visible wavelengths. 
High resolution astronomical images of
cometary comae revealed jets and other structures in the inner
parts, some of which formed spirals which rotated like water
from a lawn sprinkler indicating discrete dust emissions from
localised active parts of the nucleus.
A consequence of observing in visible light is that the results are biased
to particle sizes in the range of 1 to 10 \micron, because much smaller and much
larger particles do not contribute significantly to the scattered light \cite{gruen-cometsI}. 
With the extension of the observable spectral range to infrared wavelengths, 
also the thermal emission of dust became
accessible to astronomers. It revealed information on the abundance of larger
grains and on the mineralogical composition of the dust, the latter from
characteristic spectral features in the near and mid-infrared range.

Breakthroughs in understanding cometary constituents came
with the space missions to several comets: Giotto and two VeGa
spacecraft to comet 1P/Halley in 1986, Deep Space 1 to comet
19P/Borelly in 1999, Stardust to comet 81P/Wild 2 in 2004, and
Deep Impact to comet 9P/Tempel 1 in 2005. Water and CO were
identified as the main species in the gas, and dust particles
made of carbonaceous and silicate materials ranging from
nanometre to millimetre sizes were detected. Active areas and corresponding
dust jets were identified in spatially 
resolved images of some of the comets visited by spacecraft
\cite{keller-delamere1987,thomas-keller1987,soderblom-boice2004,sekanina-brownlee2004}.
For 67P/C--G, however, such
detail will only be observed when in 2014
the Rosetta spacecraft reaches the comet. Until then any information on the
dust environment has to be derived from astronomical observations of
the target comet or by assuming correspondence to other, better studied
comets.

The purpose of the present paper is twofold: on the one hand, to give an
overview of the
current knowledge of the dust environment of comet 67P/C--G, and on the other
hand, to provide estimates of such quantities as the spatial density, flux,
and speed of dust in the coma as functions of location and
time. These values are meant to support the planning of measurements 
of the instruments on board Rosetta. 
Section~\ref{sec:obs} contains an overview of the available astronomical
observations of comet 67P/C--G, and measurements of the albedo and temperature
of dust from \cg\ are presented. In addition, the phase function, size
distribution and radiation pressure efficiency of cometary dust are defined
and discussed. In Section~\ref{sec:model}, several methods are presented to
derive dust properties from modelling of images of the cometary tail or trail,
and results for \cg\ that were obtained by several authors are
discussed. Finally, 
in Section~\ref{sec:mmmodel}, the ESA Cometary Dust Environment Model
\cite{mueller1997, mueller1998, muellerPhD, landgraf-mueller1999} is briefly
described and results from applying it to
comet 67P/C--G are presented.

\section{Observations of \cg\ Dust and Dust Properties}
\label{sec:obs}

In this section, we give an overview of the available observational data
containing information on the dust of \cg, and we introduce and discuss the
major quantities that can be measured by means of such observations. Some of
these quantities (the dust size distribution, the radiation pressure
efficiency and the emission speeds) can -- in the absence of in situ
measurements -- only be inferred through modelling of
astronomical images, which is discussed in Section~\ref{sec:model}. 
Published astronomical data on the dust of \cg\ include images of the dust
coma, tail and trail in both visible and infrared light. They 
are available from 1982 onwards. 

\subsection{Disambiguation: Dust tail, antitail, neckline, and trail}
\label{subsec:dust_phenom}
%
In the following we give the definitions of some observational dust phenomena
associated with comets as they are used in this paper:

%
%
Outside the inner coma -- in which the dust is accelerated by gas drag --
dust dynamics is dominated by solar gravity and radiation
pressure. Both forces follow a $1/r_{\rm h}^2\,$-law ($r_{\rm h}$ being the
heliocentric distance) but act in opposite directions. Consequently, their ratio $\beta$ depends only on the material
properties of the dust grains, such as size, composition, density, and
shape. In general, radiation pressure is most efficient for particles of about
the size of the dominant wavelength of the radiation.
Such grains are driven away from the nucleus
in the direction opposed to the Sun and trailing the nucleus, thus forming the
comet's {\em dust tail} before they disperse into interplanetary space. 

The term {\em antitail} refers to a part of the tail that seems to point
toward the Sun instead of away from it. Often, this is a projection effect
that occurs when the observer is in such a position that part of the normal
tail appears to be on the Sun-facing side of the nucleus. Viewed
in three dimensions, there is no difference between an ordinary tail and a
projection antitail, but the dust seen in the antitail tends to be the larger
and older component of the dust in the tail.

A {\em neckline} \cite{kimura-liu1977, richter-curdt1991, mueller-green2001} consists of 
large particles emitted at a true anomaly of 180$^\circ$ before the
observation: 
The orbital periods of large particles are similar to that of the
parent comet. 
Their orbits are generally inclined with respect to the comet orbit, but the
particles cross the orbital plane of the comet twice during each revolution
around the Sun. One intersection point is the point of
emission. The other lies on the line of nodes connecting 
the emission point and the Sun. The position of the second intersection point
on the line of nodes depends on the emission velocity and $\beta$ of the particle.
Large particles emitted at a given time 
cross the orbital plane of the comet almost simultaneously, but at
different positions along the nodal line. To an observer in -- or close to --
the comet orbital plane, they appear as a bright line, the neckline.
Necklines can appear both in the Sun- and the anti-Sun direction i.e. can
contribute to the tail as well as to the antitail. In the case of 
comet C1995 O1 Hale-Bopp in early 1997 the neckline was mainly visible along
the tail direction, but also gave rise to an antitail \cite{boehnhardt2003}.

The {\em dust trail} of a comet consists of mm- to cm-sized particles that
-- because of low emission speeds and little sensitivity  to radiation
pressure --
remain close to the comet orbit for many revolutions around the Sun and whose
appearance reminds of an airplane contrail. Trails of eight short-period comets were first
observed with IRAS in 1983
\cite{sykes-lebofsky1986,sykes-hunten1986,sykes-walker1992a}, one of them
being that of 67P/C--G.

\subsection{Morphology of Coma, Tail, Antitail and Trail}
\label{subsec:obs_morph}

\begin{figure}[htp]
\includegraphics[clip,width=.5\textwidth]{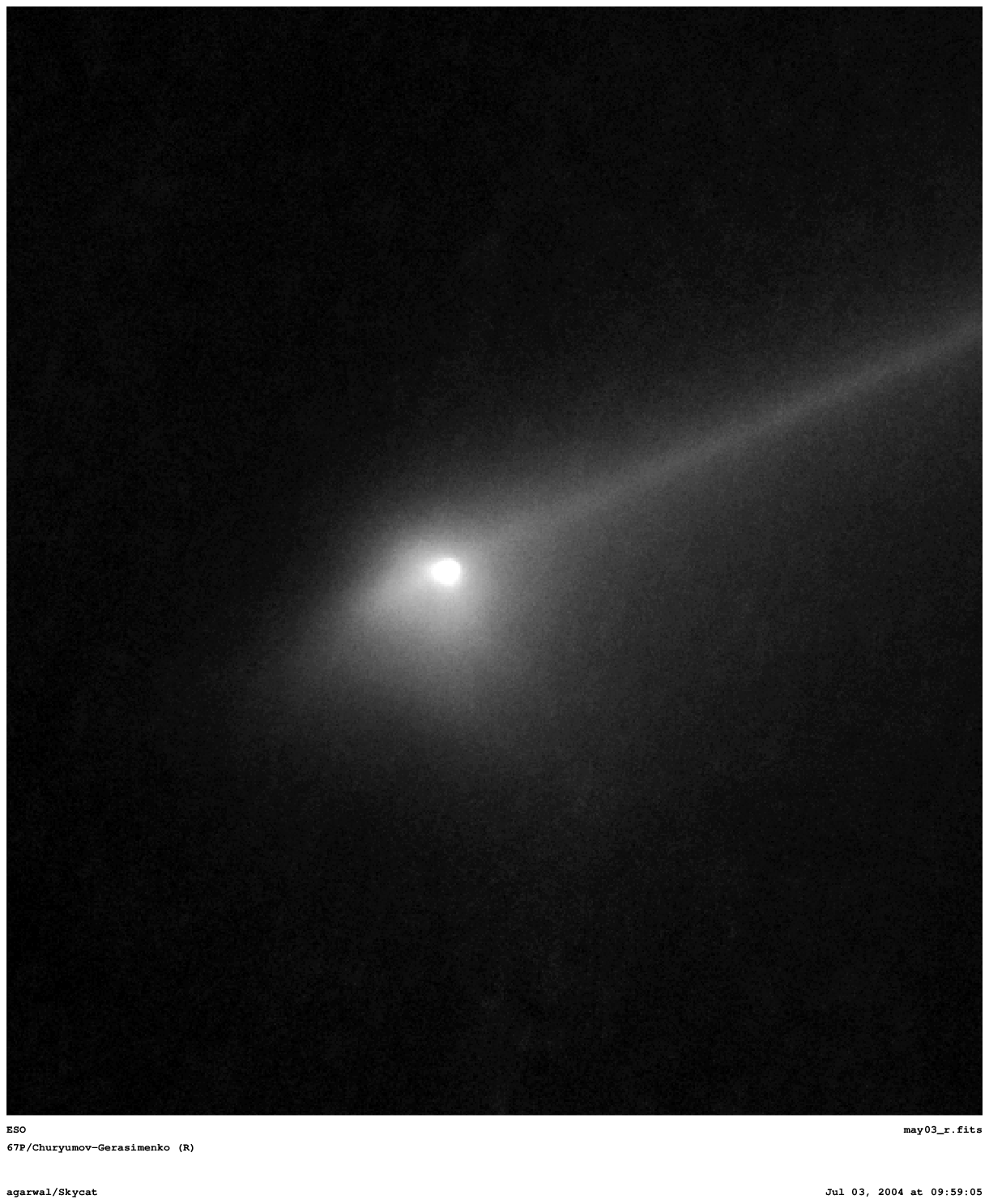}
\includegraphics[width=.5\textwidth]{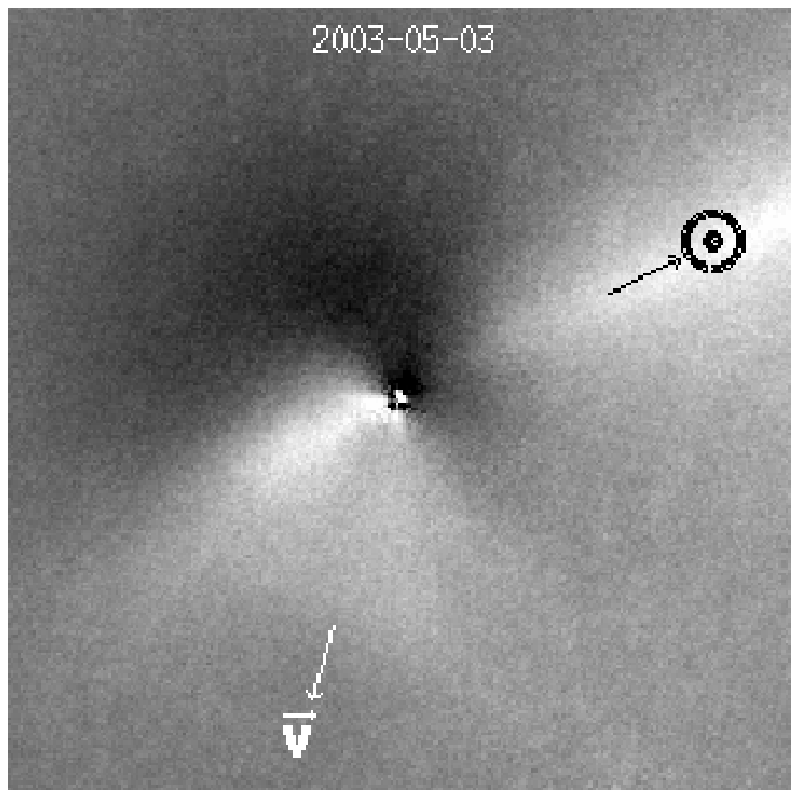}
\caption{Left: Broadband R image of comet 67P/C--G on 3 May 2003 obtained with
  FORS1 at ESO/VLT/U1 \cite{schulz-stuewe2004a, schulz-stuewe2004b}. Right:
  Same image, but structurally enhanced by subtracting the azimuthally
  averaged coma profile from the measured brightness distribution. Distinct
  features in the coma are clearly visible. The image
  is a reproduction of Figure 1 (lower left) in \cite{schulz-stuewe2004b},
  courtesy by R. Schulz. 
}
\label{fig:dust_jets}
\end{figure}

The coma of 67P/C--G showed azimuthal brightness variations during both the
  1996/97 and the 2002/03 apparitions \cite{schleicher2006, schulz-stuewe2004a,
  schulz-stuewe2004b, weiler-rauer2004a}. An example is displayed in Figure~\ref{fig:dust_jets}.
The bright regions have been interpreted as border lines of coma fans produced
  by active areas at different latitudes on the rotating nucleus. The 2003
fan pattern suggests the presence of 2 or 3 active regions on the nucleus.
For the 2-active-region scenario, the rotation axis requires to be very much
inclined to the orbital plane, while for the 3-active-region scenario a very
wide range of rotation axis directions is possible (H.~B{\"o}hnhardt, private
communication). The azimuthal direction of the bright features has been used
  to constrain the orientation of the rotation axis of the nucleus
  \cite{weiler-rauer2004a,schleicher2006}.  

The tail of 67P/C--G was characterised by a thin, bright feature close to the projected comet orbit
and pointing away from the nucleus in the direction opposed to the motion of
the comet. This feature was first observed shortly after perihelion in August
2002 and prevailed at least until April 2006 when the comet had already passed
aphelion \cite{fulle-barbieri2004a, moreno-lara2004, agarwal-boehnhardt2007,
  agarwalPhD, ishiguro2008}. Different interpretations of this phenomenon are discussed in
Section~\ref{sec:model} with the conclusion that it most probably was a
very pronounced antitail due to the low inclination of the comet orbit with
respect to the ecliptic.

The dust trail of 67P/C--G was first observed 
with the Infrared Astronomical Satellite (IRAS) in 1983
\cite{sykes-lebofsky1986,sykes-hunten1986,sykes-walker1992a} with a 
reported length of 1.2$^\circ$ in mean anomaly and a width of 50000 km. 
In visible light, observations of the 67P/C--G trail 
were done in 2002/03 at heliocentric distances between 1.3 AU and 3.1 AU
\cite{ishiguro2008} and in 2004 at $r_{\rm h} = 4.7 \,{\rm AU}$
(out-bound) \cite{agarwal-boehnhardt2007}. 
Infrared observations of the 67P/C--G trail between 2004 and 2006 were
obtained with the MIPS instrument of NASA's Spitzer Space Telescope at 24$\mu$m
\cite{kelley-reach2008, agarwalPhD}.  

\begin{figure}[htp]
\includegraphics[width=\textwidth]{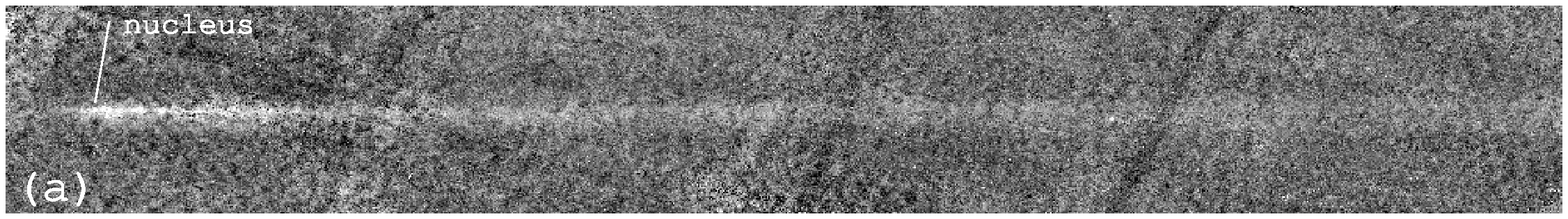}
\includegraphics[angle=270,width=\textwidth]{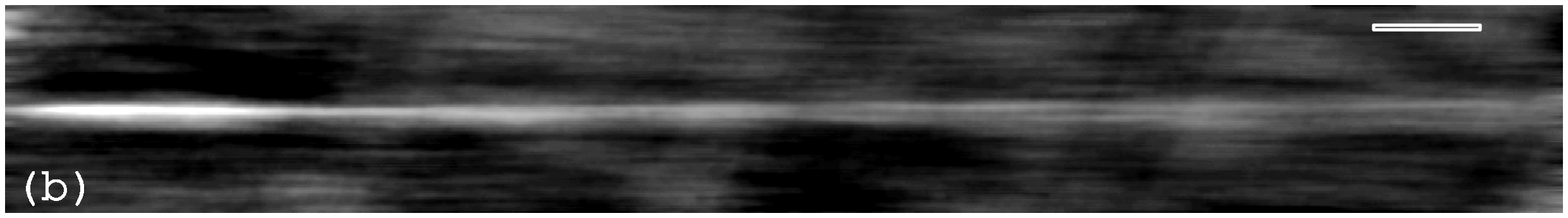}
\caption{Reproduction of Figure 1 in \cite{agarwal-boehnhardt2007}. The images
  show the dust trail and neckline of 67P/C--G in April 2004. The data were
  obtained with the Wide Field Imager (WFI) at the ESO/MPG 2.2m telescope in
  La Silla with a total exposure time of 7.5 hours and without filter. The
  size of the images is 35$^{\prime}$\,x\,4.7$^{\prime}$ each, corresponding
  to $1.1^\circ$ in mean anomaly parallel to the orbit. 
  (a) Unfiltered image. 
  (b) Same image, each pixel being replaced by the average over a
  neighbourhood of 200 pixels (140$^{\prime\prime}$)
  parallel and 10 pixels (7$^{\prime\prime}$) perpendicular to the trail
  axis after removal of the nucleus. The
  filtering window is indicated in the upper right corner of (b). 
  A detailed discussion of the data acquisition, processing and
  interpretation is given in \cite{agarwal-boehnhardt2007} and modelling
  results are given in Section~\ref{subsec:model_trail}.}
\label{fig:trail}
\end{figure}

\subsection{Albedo and Phase Function}
\label{subsec:alb_phfn}
The geometric albedo $p$ of an object is defined as the ratio of the intensities reflected
backwards by the object and by a totally diffusely
reflecting (i.\,e. Lambert scattering) disc of the same geometric cross
section \cite{hanner-giese1981}, while the Bond albedo $A_{\rm B}$ is the ratio of the
total light reflected from a sphere to total light incident on it
\cite{allen1973}. 
The phase
angle $\alpha$ is the angle between the directions of the observer
and of the incident radiation as seen from the scattering particle.  
The phase function $j(\alpha)$ describes the
ratio of intensity scattered in the $\alpha$-direction to the intensity
scattered at $\alpha = 0$, and the integral of the phase function over 4$\pi$
solid angle is called the {\em phase integral}, $q$. The Bond and geometric
albedos are related by $A_{\rm B} = q\,p$.

\begin{figure}[h]
\centering
\includegraphics[clip,width=.7\textwidth]{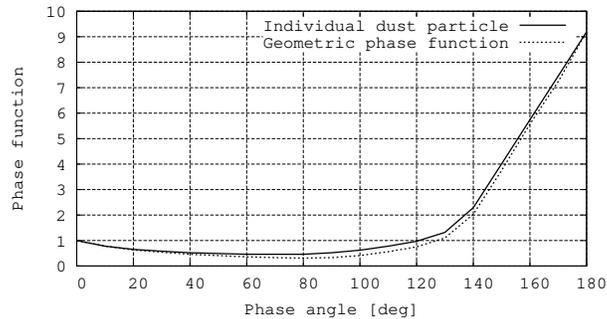}
\caption{Solid line: phase function of an individual dust particle as given in
  \cite{divine1981}, but here normalised to $j (\alpha\! = \!0) = 1$. Dashed
  line: 
  geometric phase function $j_{\rm geo} (\alpha)$ \cite{muellerPhD} accounting
  for the anisotropy of an axis-symmetric coma with peak activity at the
  subsolar point.} 
\label{fig:phase_fns}
\end{figure}

To derive the dust phase function from astronomical observations, an
object must be observed at different phase angles, i.\,e. at
different times. Since the total cross section of dust is not constant with
time, an appropriate normalisation is required, for which two methods are
used.
One method employs gas production rates measured
simultaneously with \afr\ for normalisation
\cite{millis-ahearn1982,meech-jewitt1987,schleicher-millis1998}, assuming that
the dust-to-gas ratio as well as the dust size distribution and material
properties remain constant over time. 
The other
method, preferable but more laborious, normalises the scattered intensity
to the simultaneously measured thermal infrared emission from the same
volume
\cite{tokunaga-golisch1986,hanner-tedesco1985a,hanner-newburn1989,gehrz-ney1992,ney1974,ney-merrill1976,ney1982}.
The general shape of the phase function of cometary dust is characterised by a
distinct forward and a gentle backscattering peak and is rather flat at medium
phase angles \cite{kolokolova-hanner-cometsII-2004}. Divine \cite{divine1981}
derived from data given in \cite{ney1974,ney-merrill1976,ney1982} the phase
function shown in Figure~\ref{fig:phase_fns} (solid line). The figure also
shows the geometric phase function that describes the phase angle dependence
of \afr\ and is discussed in Section~\ref{subsec:obs_afrhogas}.

Laboratory
measurements and theoretical studies suggest that the dust albedo depends on
particle size \cite{mcdonnell-lamy1991}. Earth based observations have so far
not been suitable to investigate this dependence, because coma observations 
only provide data for the ensemble of particles of all sizes along a line of
sight. 

The geometric albedo is derived from the simultaneous observation of
the scattered visible and the thermally emitted infrared light, either directly at $\alpha =
0$, or at multiple phase angles and assuming a given phase
function. 
The geometric albedo of dust in the coma of \cg\ derived from the optical and infrared
brightness was 0.04 at 1.25\,\micron\ and 0.05 at 2.2\,\micron\
\cite{hanner-tedesco1985a}, which is in accordance with a large sample of
comets \cite{divine-fechtig1986,hanner-newburn1989}. 
From the low
 albedo it is inferred that
there is no significant population of cold, bright (and possibly icy) grains that would
contribute to the scattered light but not to the thermal emission \cite{hanner-tedesco1985a}. There is,
however, some indication that the geometric albedo is higher for comets beyond
$3 \,{\rm AU}$ \cite{hanner-newburn1989}, which may point in the same
direction. 

\subsection{Afrho and Gas Production}
\label{subsec:obs_afrhogas}

The brightness of a cometary coma is proportional to the dust production rate.
To infer the production rate from data obtained under different observational
circumstances, the measured brightness must be corrected for all other 
parameters on which it depends.
The quantity \afr\ was defined for this purpose \cite{ahearn-schleicher1984}. \afr\ stands for the
product of albedo $A = p j(\alpha)$ (see Section~\ref{subsec:alb_phfn}), 
filling factor $f$ of grains within the field of view,
and the radius $\rho$ of the aperture at the comet. It is measured
as follows:
\begin{equation}
\label{eq:afr-def}
Af\!\rho=4\frac{\Delta^2 (r_{\rm h} / 1 {\rm AU})^2} 
{I_{\rm sun}^{\rm (filter)}} \times
\frac{I_{\rm dust}^{\rm (filter)}}{\rho},
\end{equation}
where $r_{\rm h}$ and $\Delta$ are the heliocentric and geocentric
distances of the comet during the observation, and $\rho$ is the
radius of the circular aperture on which the coma intensity $I_{\rm dust}^{\rm
  (filter)}$ was measured using a given filter. $I_{\rm sun}^{\rm (filter)}$ is the intensity of the Sun at 1\,AU heliocentric distance seen through the same filter.
Provided that the dust particles move away from the nucleus on 
straight trajectories and are not subject to
processes altering their scattering behaviour, \afr\ is independent of
the employed aperture radius, of the heliocentric and geocentric distances, 
and -- to the extent that the dust can be considered as ``grey''
-- of the spectral band in which the observation was carried out.

For an isotropic coma and discrete dust sizes $s_j$, \afr\ is related to the
production rates $Q_{{\rm d},j} \,(s_j)$ via the dust emission speeds $v_{{\rm
    d},j}$,
the geometric albedo $p$, and the phase function $j(\alpha)$ \cite{muellerPhD}:   
\begin{equation}
\label{eq:afrnaper}
Af\!\rho=2\pi \, p \, j(\alpha) \sum_{j} s_j^2 \,\frac{Q_{{\rm d},j}}{v_{{\rm d},j}}.
\end{equation}
The relative magnitudes of the $Q_{{\rm d},j}$ rates are given by the size
distribution.

\afr\ depends on the phase angle of the observation, due both to the
scattering properties of a single dust grain \cite{schleicher-millis1998} and
-- unless the coma is isotropic -- to projection effects \cite{muellerPhD}. 
In a non-isotropic coma, the timescale on which the particles leave a given
field 
of view depends on the angle between the main emission direction and the line
of sight, with higher measured \afr\ for a line of sight
parallel or close to the main emission direction. 
The phase-angle dependence of \afr\ is then better described by a 
{\em geometric} phase function $j_{\rm geo} 
(\alpha)$ \cite{muellerPhD} than by that of a single particle.
The geometric phase function is characteristic of the specific pattern of
emission of a given nucleus. 

In practice, \afr\ often depends on the aperture size despite its
definition \cite[and references therein]{schleicher-millis1998,schleicher2006}. This implies that the brightness distribution in the coma deviates
from the assumed 1/$\rho$-profile.
Possible causes for this deviation include
changes in the physical properties of the grains as they travel
outward (e.\,g. loss of volatiles or fragmentation), the action of
radiation pressure modifying the straight trajectories of small particles
inside the field of view,
or a long-lasting 
population of large particles \cite{schleicher-millis1998}.

\begin{figure}[hbp]
\includegraphics[width=.5\textwidth]{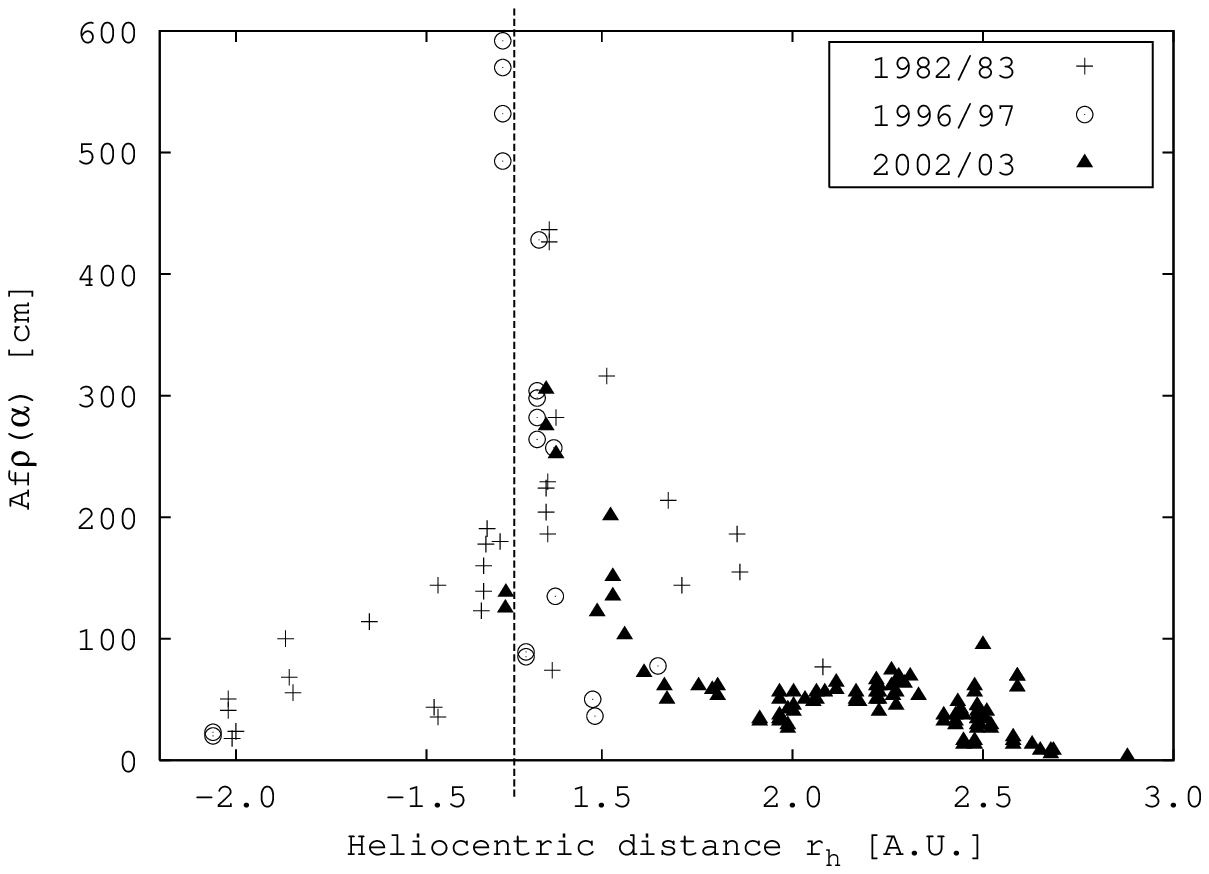}
%
\includegraphics[width=.5\textwidth]{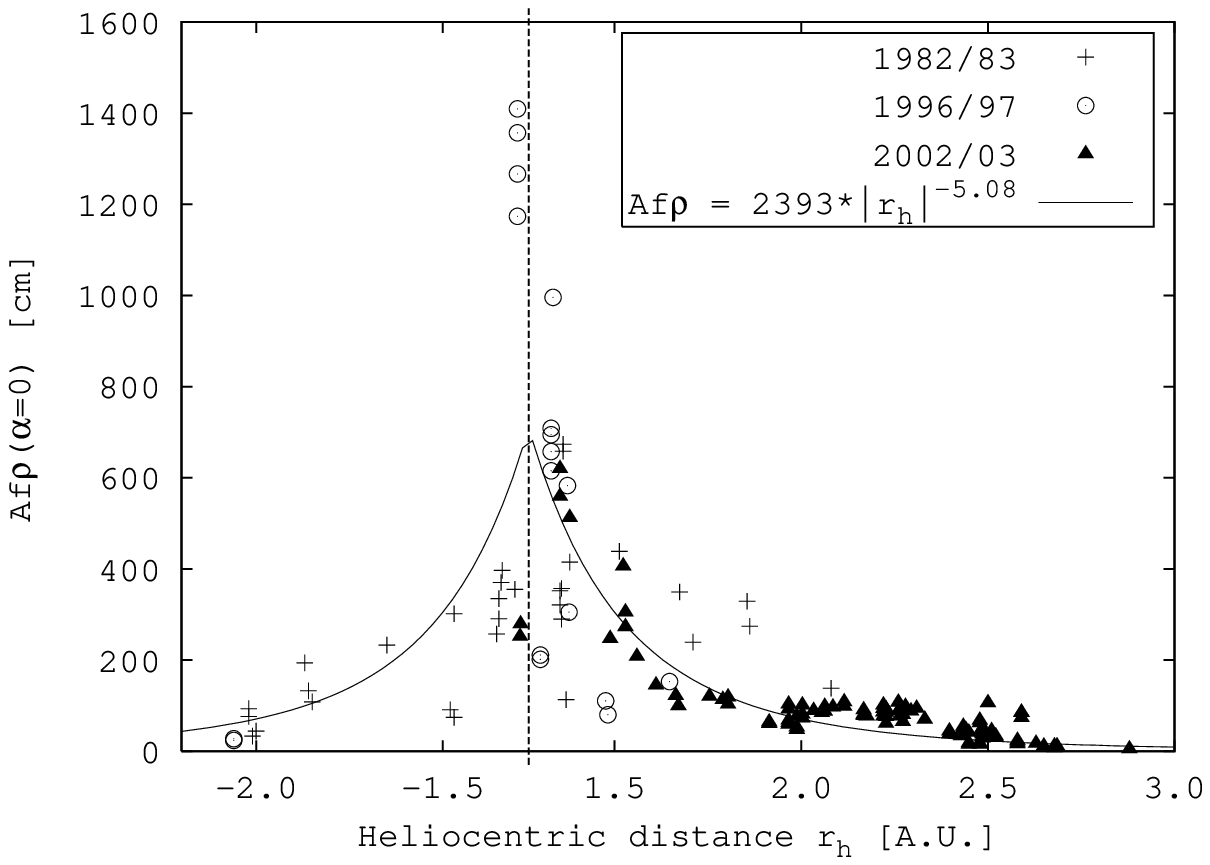}
%
\begin{center}
\includegraphics[width=.5\textwidth]{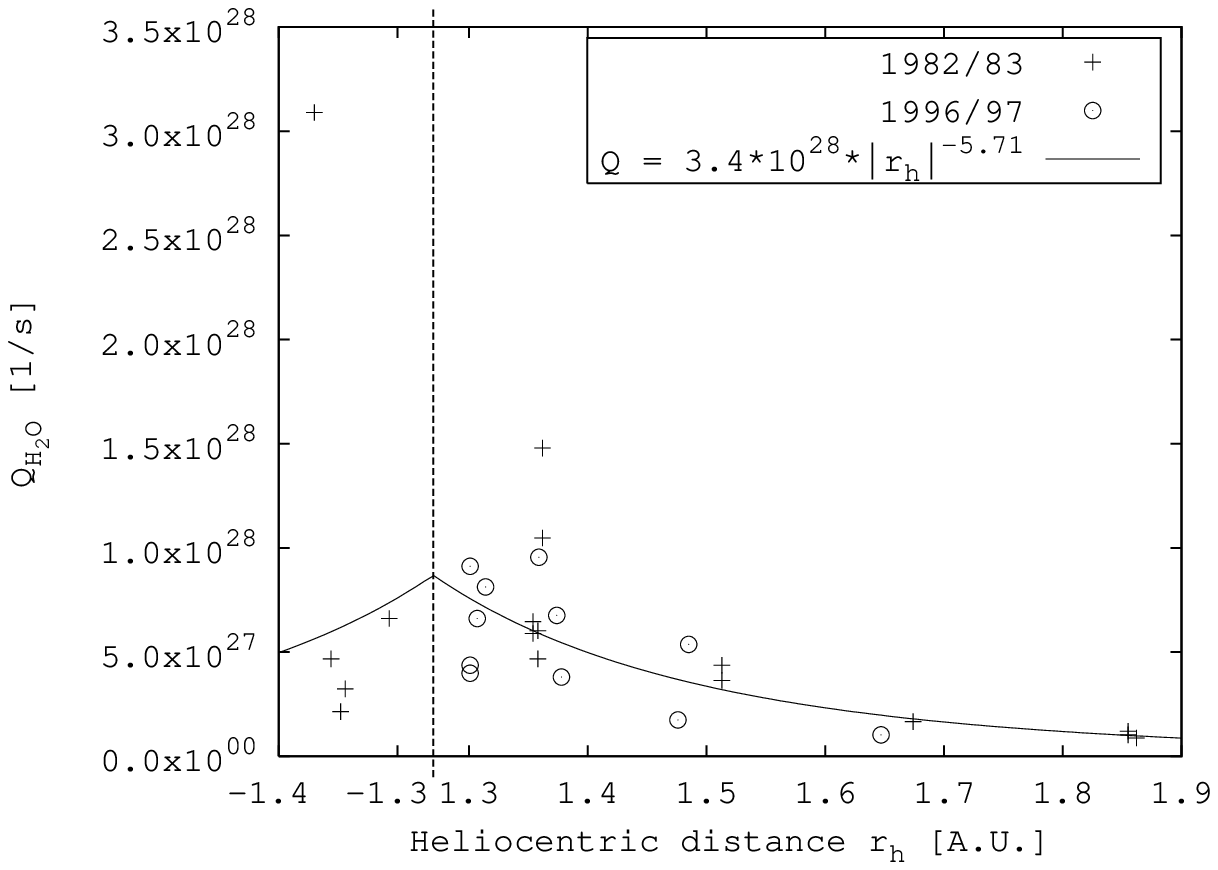}
\end{center}
\caption{Top left: observed \afr~values as function of heliocentric distance
  $r_{\rm h}$. Top right: observed \afr~values corrected for phase angle
  through division by the geometric phase function (Figure~\ref{fig:phase_fns}). Power-law fit for phase angle corrected \afr: $Af\!\rho\,
 (r_{\rm h}, \alpha\!\!=\!\!0$) $=2393\,\times (r_{\rm h} / 1 \, {\rm
  AU})^{-5.08}$ cm. Bottom: Measured H$_2$O production rates with
  corresponding power-law fit $Q_{\rm H_2O} \,(r_{\rm h}) = 3.4
  \times\!10^{28}\,\times ( r_{\rm h}/ 1 \, {\rm AU})^{-5.71}$ molecules/s. }
\label{fig:afr_h2o}
\end{figure}

\afr\ has been measured for 67P/C--G during three perihelion
passages (1982/83, 1996/97 and 2002/03) \cite{storrs-cochran1992a,osip-schleicher1992a,ahearn-millis1995a,kidger2003,lamy2003a,weiler-rauer2004a,schulz-stuewe2004a,feldman-ahearn2004,schleicher2006}\footnote{The
  data from these publications are summarised in
  the data base at  
  \mbox{http://berlinadmin.dlr.de/Missions/corot/caesp/comet{\_}db.shtml}. The
  site also includes data obtained by amateur astronomers.}.
Figure~\ref{fig:afr_h2o}\,(top left) shows all published \afr\
data as function of heliocentric distance.
The same values after correction for the phase angle dependence using the
geometric phase function (Figure~\ref{fig:phase_fns}) are displayed in
Figure~\ref{fig:afr_h2o}\,(top right) together with a power-law fit to the
corrected values. 
Figure~\ref{fig:afr_h2o}\,(bottom) shows observed production rates of H$_2$O
\cite{hanner-tedesco1985a,feldman-ahearn2004,maekinen2004,crovisier-colom2002a}
and a power-law fit to the data. The exponents of the derived
power laws are untypically steep compared with other comets, and more data of
high quality will be needed to confirm them.
Both \afr\ and the H$_2$O production rate reach their maxima around 30 days
after perihelion. No dust coma was detected 
beyond at least 4.9 AU \cite{tubiana-drahus2007_dps}. 

\subsection{Dust Temperature}
\label{subsec:obs_temp}
Assuming that a particle is characterised by the Bond albedo $A_{\rm B}$ at visible
wavelengths and the emissivity $\epsilon$ in the
infrared, its temperature $T$ at the heliocentric distance $r_{\rm h}$ (in AU) is
given by the equilibrium between absorbed solar and emitted thermal radiation:
\begin{equation}
\label{eq:gbeq}
\frac{(1\!-\!A_{\rm B}) \, \Isun}{r_{\rm h}^2} = 4\, \epsilon \sigma \,T^4,
\end{equation}
where $\sigma$ is the Stefan-Boltzmann constant
and $\Isun$ = (1367 $\pm$ 2) W m$^{-2}$ the solar flux at 1\,AU \cite{cox2000}. The resulting equilibrium temperature is
\begin{equation}
T (r_{\rm h}, A_{\rm B},\epsilon)
= 278.8 \,{\rm K} \, \left(\frac{1-A_{\rm B}}{\epsilon}\right)^{\!\frac{1}{4}} \, \frac{1}{\sqrt{r_{\rm h}}}.
\label{eq:T_eq}
\end{equation}
A blackbody would be characterised by $A_{\rm B}$ = 0 and $\epsilon$ = 1.
The temperatures of dust in the inner solar system are generally such that the
main emission lies in the infrared.
In practice, the temperature
is derived from fitting a blackbody spectrum to
measurements of the brightness at different infrared wavelengths.

The coma of \cg\ was monitored at multiple wavelengths in the range of 1 to
20\,\micron\ between September 1982 (1.50 AU preperihelion) and March 1983
(1.88 AU postperihelion) \cite{hanner-tedesco1985a}. The derived 
temperatures of the dust in the coma 
were throughout higher than those of a theoretical blackbody
at the same heliocentric distance, which is generally attributed either to the
presence of submicron-sized particles (i.e. smaller than the dominant
wavelength range of the thermal emission) \cite{hanner-tedesco1985a} or
to very porous aggregates of small grains \cite{hanner-newburn1989,
kolokolova-hanner-cometsII-2004}.
For a given heliocentric distance,
the colour temperature was higher post- than pre-perihelion,
suggesting a change in the particle properties or in the dominant size. 
Additionally, an 8- 13-\micron\ spectrum was taken on 23 October 1983. It did
not show a silicate feature, which is usually
taken as an indication for the dominance of somewhat larger and more 
compact grains. 

Excess colour temperatures were also derived from IRAS observations of
come\-tary trails at 12, 25, and 60\,\micron\ \cite{sykes-walker1992a}.
For \cg, the derived temperature at 2.3 AU was approximately 14\%
above that of a blackbody at the same heliocentric distance, 
corresponding to $\epsilon/(1-A_{\rm B})$ = 0.6 $\pm$ 0.2.

\begin{figure}[htp]
\centering
\includegraphics[width=.7\textwidth]{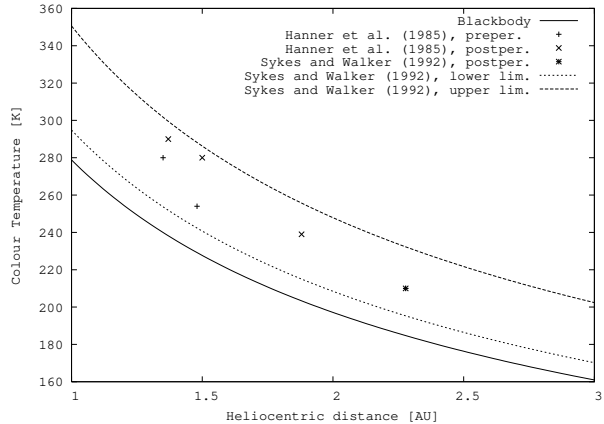}
\caption{Colour temperatures of the dust from \cg\ derived from
  multi-wavelength infrared observations. The symbols ``+'' and ``x'' refer to
  observations of the coma before and after perihelion, respectively
  \cite{hanner-tedesco1985a}. The asterisk indicates the temperature of
  the dust trail \cite{sykes-walker1992a}, and the two dashed lines correspond
  -- via Equation~\ref{eq:T_eq} --
  to the upper and lower limit of $\epsilon/(1-A_{\rm B})$ derived from the same data.
  The solid line shows the temperature of a theoretical blackbody, described
  by Equation~\ref{eq:T_eq} with $A_{\rm B}$ = 0 and $\epsilon$ = 1.
}
\label{fig:dust_temp}
\end{figure}

Figure~\ref{fig:dust_temp} shows the temperature measurements discussed above
and -- for comparison -- the equilibrium temperature of a blackbody as a function of heliocentric
distance. The data from the coma \cite{hanner-tedesco1985a} and from the trail 
\cite{sykes-walker1992a} are remarkably consistent given that coma and trail
are generally assumed to be dominated by different particle populations
(\micron- versus mm-cm sized).

\subsection{Dust Size Distribution}
\label{subsec:size_distr}
The size distribution of cometary dust has
been inferred from both astronomical images and in situ data. 
While the former yield a {\em size} distribution, the latter contain
information on the {\em masses} of the particles.
For optical images, the determined sizes scale directly with
the particle albedo. The conversion
from size to mass requires knowledge of the  bulk density of the particles.

A mass or size distribution can be specified in the form of either a {\em 
differential} or a {\em cumulative} distribution. The cumulative mass
distribution $F(m_0)$ gives information on the fraction $N$ of
particles that have a mass greater than some mass $m_0$:
\begin{equation}
N \, (m > m_0) = F(m_0). 
\end{equation}
The differential mass distribution $f(m)$ characterises the
relative abundance $n$ of particles inside a mass interval $[m_1,m_2]$:
\begin{equation}
n \, (m_1 \!<\! m \!<\! m_2) = \int\limits_{m_1}^{m_2} f(m) \: \df m
= F(m_1) - F(m_2).
\label{eq:nfrac_from_md}
\end{equation}
If the mass of a particle can be converted to a size by a relation
$s(m)$, a corresponding differential size distribution $g(s)$ exists:
\begin{equation}
\int\limits_{m_1}^{m_2} f(m) \: \df m = \int\limits_{s(m_1\!)}^{s(m_2\!)}
g\,(s) \: \df s.
\end{equation}
It is generally assumed that -- at least in intervals of the total mass
range covered by cometary dust -- the
distributions can be approximated by power laws.
In the literature, both the
exponents $\gamma$ of the {\em cumulative mass distribution} $F(m) \sim
m^{-\gamma}$ and $\alpha$ of the {\em differential size distribution} $g(s)
\sim s^{\alpha}$ are commonly used. For a constant 
bulk density in the concerned size interval we have $m(s) \propto s^3$
and $\df m \propto s^2 \df s$. Hence the exponents
$\gamma$ and $\alpha$ are related by
\begin{equation}
\alpha = -3 \gamma -1.
\label{eq:alpha2gamma}
\end{equation}
The mass 
distribution at the nucleus is different from the one in the coma because of
the size-dependence of the emission velocities.
In 
general, large particles are more abundant in the coma than close to the
surface because of their lower speeds.
The relation between the size distributions in the coma and at the
nucleus may be additionally complicated by fragmentation or evaporation of
grains \cite{mcdonnell-alexander1987,mcdonnell-lamy1991,green-mcdonnell2004} and by an
inhomogeneous distribution of surface activity \cite{fulle-colangeli1995}. 

In situ data on the dust mass distribution were obtained by the dust
instruments on board the
spacecraft VeGa 1 and 2 and Giotto at comet 1P/Halley in 1986
\cite{mcdonnell-alexander1987,divine-newburn1987b,mcdonnell-lamy1991,fulle-colangeli1995}, on board Giotto at comet 26P/Grigg-Skjellerup in 1992
\cite{mcdonnell-mcbride1993}, and on board Stardust at comet 81P/Wild~2 in 2004
\cite{tuzzolino-economou2004, green-mcdonnell2004}. 
The measured quantity is not 
the mass distribution of dust as released from the nucleus but the cumulative
flux or fluence on the concerned instrument. 
The fluence is defined as the flux integrated over the spacecraft trajectory,
it represents therefore an average mass distribution. 
The flux (or the
fluence measured in only sections of the trajectory) showed significant
variation with time during both the 1P/Halley and the 81P/Wild~2 fly-bys.

Table~\ref{tab:md_lit} lists mass- or size-distribution exponents given in
the literature. 
Still no general agreement has been reached on the interpretation of
the data with respect to the dust mass distribution at the nucleus, but the
authors listed in Table~\ref{tab:md_lit} do agree that the mass of dust in the
coma is dominated by the largest emitted  
particles. The cumulative fluences registered at the various
spacecraft show different exponents for large and small particles. The
interpretations of this observation are not unanimous, either.

\begin{table}[h]
\caption{Exponents of the cumulative fluence on a spacecraft, $\tilde{\gamma}$, the cumulative
  mass distribution, $\gamma$, and the differential size distribution,
  $\alpha$, 
  compiled from the literature. {\bf Bold} values are taken directly from the
  publications, the remaining values were derived assuming the following relations.
 To translate $\tilde{\gamma}$ to $\gamma$, it is assumed that
  the speeds of small particles are limited by the speed of the gas and
  therefore independent of size. Hence $\gamma = \tilde{\gamma}$ for small
  particles  (indicated by ``$\dagger$''). For larger particles, the relation
  $v \propto m^{-1/6}$ is used, which was derived from a radially symmetric
  coma model \cite{divine-fechtig1986}, resulting in $\gamma = \tilde{\gamma}
  + 1/6$ for large particles (indicated by ``$\ast$''). The relation between
  the differential size distribution and the cumulative mass distribution  
is given by $\alpha = -3\gamma - 1$. Results from modelling of astronomical
  images are not included here. For \cg\, a compilation of size distributions
  derived by various authors is shown in Figure~\ref{fig:comparison_model}\,(bottom).}
\begin{center}
\label{tab:md_lit}
\begin{tabular}{@{\extracolsep{\fill}}llr@{\,\ldots\,}lllll}
\hline\noalign{\smallskip}%
Comet                 & Instrument(s) &\multicolumn{2}{c}{Mass/size interval}& $\tilde{\gamma}$ & $\gamma$      & $\alpha$ & Ref.\\
\hline\hline\noalign{\smallskip}
1P/Halley             & DIDSY\&PIA    & $10^{-13}$ & $10^{-8}$ kg           &                  & {\bf 1.02}     & $-$4.06    & \cite{mcdonnell-alexander1987}\\
                      &               & $10^{-8}$  & $10^{-6}$ kg           &                  & {\bf 0.71}     & $-$3.13    &\\
                      &DIDSY\&PIA     & $10^{-19}$ & $2\!\times\!10^{-14}$ kg   & {\bf 0.18}       & $0.18^\dagger$ & $-$1.54    &\cite{divine-newburn1987b}\\
                      &               & $2\!\times\!10^{-14}$ & $10^{-6}$ kg    & {\bf 0.94}       & $1.11^\ast$    & $-$4.32    &\\

                      &VeGa 1         & $10^{-19}$ & $10^{-12}$ kg          & {\bf 0.26 }      & $0.26^\dagger$ & $-$1.78&\\
                      &               & $10^{-12}$ & $10^{-9}$ kg           & {\bf 1.19 }      & $1.36^\ast$    & $-$5.07&\\ 
                      &VeGa 2         & $10^{-19}$ & $1.6\!\times\!10^{-13}$ kg & {\bf 0.26 }      & $0.26^\dagger$ & $-$1.78 &\\
                      &               & $1.6\!\times\!10^{-13}$ & $10^{-9}$ kg  & {\bf 0.90 }      & $1.07^\ast$    & $-$4.2 &\\
                      &DIDSY          & \multicolumn{2}{c}{$>$ 20 \micron}    &                  &              & {\bf $-$3.5} ${\mathbf \pm}$ {\bf 0.2} & \cite{fulle-colangeli1995}\\[4mm]
26P/G--S  &DIDSY/GRE            & $10^{-9}$ & $10^{-7}$ kg            & {\bf 0.27$^{+0.13}_{-0.20}$} & $0.44^\ast$ & $-$2.3 & \cite{mcdonnell-mcbride1993}\\[4mm]
81P/Wild 2            &DFMI           & $10^{-14}$ & $10^{-9}$ kg           & {\bf 0.85} $\pm$ {\bf 0.05} & $1.02^\ast$ & $-$4.05 $\pm$ 0.15 & \cite{green-mcdonnell2004}\\
		      &Stardust Samples& $10^{-17}$ & $10^{-3}$ kg           & {\bf 0.57 }      & $0.74^\ast$	&   $-$3.21    &\cite{hoerz-bastien2006}\\[4mm]
\noalign{\smallskip}\hline
\end{tabular}
\end{center}
\end{table}

\subsection{Radiation Pressure}
\label{subsec:rad_press}
Both solar gravity
and the radiation pressure force are inversely proportional to the heliocentric
distance squared, and point radially away from the Sun. Radiation pressure can,
therefore, be included in the equation of motion of a particle by introducing a
modified potential substituting the gravitational constant, $G$, by
$\tilde{G}$ = $G (1-\beta)$, where $\beta$ stands for 
the ratio of solar gravity to the radiation pressure force.
It depends only
on material properties of the dust grains, not on their distance from the Sun:
\begin{equation}
\beta = \frac {3 \,L_\odot}{16 \,\pi c\, G M_\odot} \frac{Q_{\rm pr}}{\rho s}.
\label{eq:beta}
\end{equation}
$L_\odot$ and $M_\odot$ are the luminosity and mass of the
Sun, and $c$ is the speed of light.
The grain has the bulk density $\rho$ and the effective
radius $s$, and it is characterised by the radiation pressure efficiency
$Q_{\rm pr}$
\cite{burns-lamy1979,divine-fechtig1986}. 
$Q_{\rm pr}$ represents the absorption and scattering properties of the grain averaged
over the solar spectrum. For homogeneous spheres and some other simple shapes, $Q_{\rm pr}$ can be
calculated in an exact way from the complex refractive index of the material
\cite{mie1908, bohren-huffman1983}. For more complicated structures, a
variety of theoretical and experimental approaches exist to obtain $Q_{\rm
  pr}$ \cite{gustafson-dustbook-2001}. 
While for particle sizes on the order of 0.01 to 1\,\micron\,,  $\beta$
depends sensitively on material, shape, structure, surface properties, and size of the particles, it is approximately constant for much smaller particles,
and proportional to 1/($\rho s$) for large ones, i.\,e. in
the geometric optics regime. This implies that for sub-millimetre and
larger particles, $Q_{\rm pr}$ is to first order independent of size \cite{burns-lamy1979,bohren-huffman1983}.

\section{Results of Modelling the 67P Dust Tail and Trail}
\label{sec:model}
In this Section, we give an overview of the efforts undertaken to
derive properties of the \cg\ dust from modelling astronomical
observations of the thin, bright feature described in
Section~\ref{subsec:obs_morph}, variantly referred to as 
neckline, antitail, or trail.
The main quantities to be constrained by the models are the dust size
distribution, the radiation pressure efficiency, and the emission speeds.
The derived values will serve as input to the hydrodynamic coma model used in
Section~\ref{sec:mmmodel} to predict the dust environment inside the coma.
In the following, we summarise the results from neckline photometry applied to
images taken in 2002/03 \cite{fulle-barbieri2004a, moreno-lara2004}, from analysis of the
antitail as observed in May 2003 \cite{agarwal-mueller2007}, and from
simulations of the dust trail in visible and infrared wide-field observations
between 2002 and 2006 \cite{agarwalPhD, ishiguro2008, kelley-reach2008}.

The position of a dust particle in a cometary tail is a function of the
radiation pressure coefficient 
$\beta$, the emission time, and the emission speed; and $\beta$ is closely related to the mass
of a particle. Hence, the effect of solar radiation
pressure is similar to that of a mass spectrometer. To understand the
formation of dust tails, the concept of synchrones and syndynes was
introduced by Bredikhin \cite{bredikhin1903}. Synchrones are the positions of
particles of different $\beta$ emitted at a given time, while syndynes describe
the positions of particles of fixed $\beta$ and varying emission
time. Both terms refer to hypothetical particles emitted with zero velocity
relative to the nucleus. Since, realistically, the initial velocity of a dust
grain is different from zero, the resulting synchrone or syndyne will have a
finite cross section with a radius proportional to the product of dust emission speed and 
time elapsed since emission. In particular, grains
released at a given time with isotropic speed will form a
spherical shell the centre of which moves along the appropriate syndyne.

This description was used by Finson and Probstein 
\cite{finson-probstein1968a,finson-probstein1968b} to derive properties of the
dust size distribution from the brightness
patterns observed in cometary dust tails. Their method is limited to recently
emitted and small particles, because it neglects tidal effects and others
(e.g. direction-dependent emission speed or production rate) that
cause a dust shell to divert from the spherical shape \cite{kimura-liu1977,
  fertig-schwehm1984, mueller-green2001}. Various approaches to surmount these
limitations are described in the following.

\subsection{Neckline Photometry}
\label{subsec:model_fulle}
Fulle et al. \cite{fulle-barbieri2004a} analyse \cg\ images
obtained in March 2003 (\rh\ = 2.6 AU) and before, when the 
comet was active and exhibited a significant coma.
They employ both an
analytical theory of neckline photometry and an inverse Monte Carlo model. 
 
The former is an analytical method to infer the emission speeds and abundance
of particles as a function of the radiation pressure parameter $\beta$
\cite{fulle-sedmak1988}. 
It is applied to an image of the comet 
obtained with the Schmidt Telescope at the Th{\"u}ringer Landessternwarte
(TLS) in Tautenburg on 27-28 March 2003.
The method relies on the assumption  
that the bright narrow feature in the tail is a neckline, and that 
all intensity observed in this feature  
is due to dust emitted within an interval of 10 hours at a true anomaly of
$180^\circ$ before that of the observation, which corresponds to 5 May
2002  (105 days before perihelion) and \mbox{\rh\ = 1.8 AU}.
The authors find that 
-- in this particular time interval -- both the mass and the cross
section of dust emitted by the comet were dominated by particles in the size
range of $1.5$ to $10 \,{\rm mm}$, and that the exponent of the differential
size distribution, $\alpha$, was between -3.5 and -3. 
From the fact that the described feature had been
constantly observable since perihelion in August 2002, they infer that 
the characteristics of the dust emission as recorded in the neckline did not
change between 3.6 and 1.7 AU before perihelion.

The second method -- the inverse Monte Carlo model -- consists in a least-squares fit of simulated images to the
measured surface intensity \cite{fulle1989a}. This method yields a set of time-dependent dust parameters
(size distribution, emission velocity, size range, \afr, and dust production
rate).
The obtained solution is unique  
in the mathematical sense of a least-squares fit; its physical probability
remains harder to evaluate.  
The method is applied to an image obtained with the Telescopio Nazionale
Galileo (TNG) on La Palma during the same night as the TLS image.
Quantitative interpretation of the model is achieved
through matching of the modelled \afr\ to the value measured during a
different observation close in time \cite{lamy2003a}. 
The authors \cite{fulle-barbieri2004a}
conclude that the time dependence of all parameters must be asymmetric with respect to perihelion
in order to match the observation. 
The size distribution exponent drops from $-3.5$ before to $-4.5$ after
perihelion, the mass loss rate from $200 \,{\rm kg/s}$ to $10 \,{\rm kg/s}$.
The emission speeds decrease by a factor of three during this time
interval (from 3\,m/s to 1\,m/s for a particle of 1\,mm radius), and the
dominant size by two orders of magnitude. 
Fulle et al. \cite{fulle-barbieri2004a} suggest the existence of two different source regions: one emitting large particles and only active before perihelion,
and one emitting mostly small particles becoming  active after perihelion.

A decreasing production rate was also found by Moreno et al. 
\cite{moreno-lara2004} applying a similar method to the same image after
calibration.

\subsection{Antitail Analysis in the Hypersonic Approximation}
\label{subsec:model_hyperson}
The dust size distribution has been evaluated  by analysis of the brightness
profile along the \cg\ tail using the Finson-Probstein model
 in the  {\em hypersonic approximation} \cite{agarwal-mueller2007}, which was
 applied to an image (Figure~\ref{fig:dust_jets}) obtained with ESO/VLT on 3
 May 2003 \cite{schulz-stuewe2004a, schulz-stuewe2004b}.

\begin{figure}[bp]
\centering
\includegraphics[width=.49\textwidth]{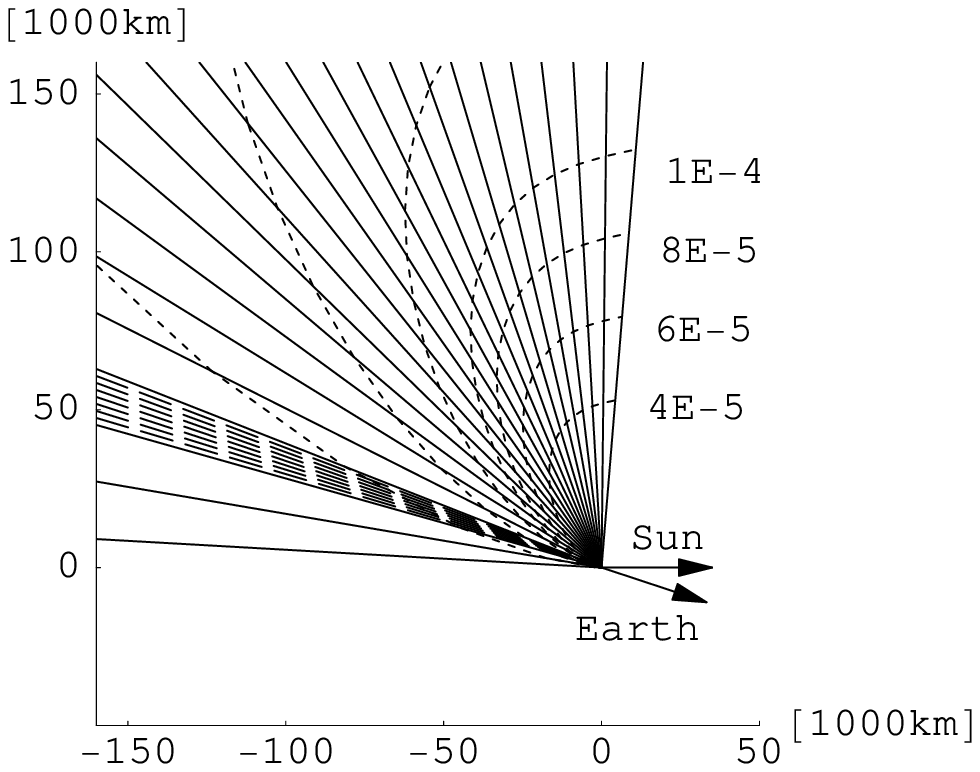}
\includegraphics[width=.49\textwidth]{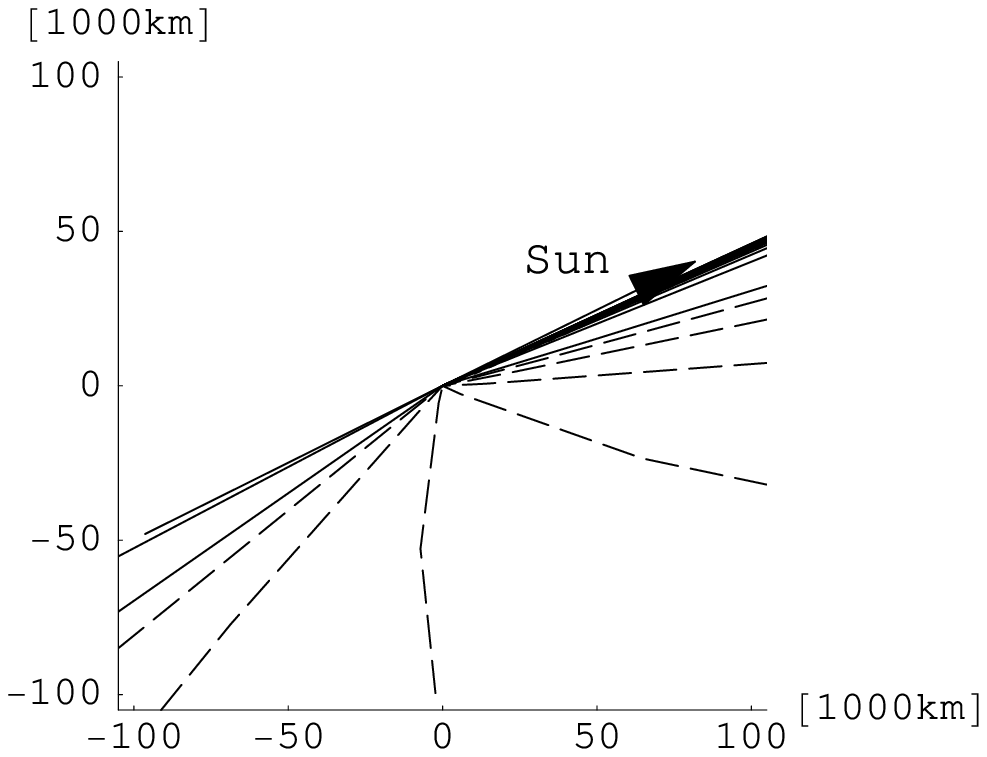}
\caption{
For 3 May 2003 synchrones are shown projected in the comet orbit plane
(top) and image plane (bottom). The solid lines correspond to synchrones
ejected in steps
of $10^\circ$ in true anomaly starting $5^\circ$ in true anomaly before the observation
time back to an ejection time that corresponds
to a heliocentric distance of 3~AU before perihelion. Synchrones corresponding
to ejection times after mid January 2003 appear in the direction opposed to
the Sun. The synchrones of particles ejected before mid December 2003 appear
in Sun direction. Only the synchrones of particles ejected in between (shown
as long dashed) have significant angles with the Sun or anti-sun direction.
Projected in the comet orbit plane (top panel) also syndynes are shown
(dashed). The $\beta$-values corresponding to the
syndynes are annotated for the larger particles. The $\beta$ values of the
smaller particles
are $1.5 \!\times\! 10^{-4}$, $3 \!\times\! 10^{-4}$, and  $1 \!\times\! 10^{-3}$.}
\label{fig:synsyn}
\end{figure}

At the time of the observation, \cg\ was at a heliocentric
distance of 2.9 AU postperihelion. A plot of synchrones and syndynes
(Figure~\ref{fig:synsyn}) shows that all synchrones corresponding to
ejection before 1.5 AU postperihelion (23 October 2002) appear projected in
the direction towards the Sun, forming an antitail. They all are -- in projection --
inclined against the neckline by less than 0.8\degree, which places doubts
upon the assumption that all dust in the bright feature was emitted during an
interval of merely 10 hours, on which the first model described in
Section~\ref{subsec:model_fulle} relies \cite{fulle-barbieri2004a}.

The hypersonic 
approximation holds when the extent of the dust shells along a synchrone
is small, such that a unique relation between the position on the synchrone
and $\beta$ can be assumed \cite{finson-probstein1968a}. 
It is then possible to infer the size distribution of the particles ejected
from the nucleus from the brightness profile along the synchrone, if one of
the following two conditions is met: (1) Either the antitail originates from
the superposition of many synchrones that -- in projection -- have only a
small offset from each other compared with the extension of the dust shells
perpendicular to the synchrone.  Or (2) the brightness along one direction in
the image plane is dominated by one synchrone alone. 
A generalisation of the original formulation of the hypersonic approximation \cite{finson-probstein1968a} --
which applied only to spherical shells -- has been developed for tidally
distorted elliptical shells, assuming radially
symmetric particle emission \cite{agarwal-mueller2007}. 


The intensity along the antitail on 3 May 2003 could be fitted by a power law
with an exponent of $-0.4$, which was translated to  $\alpha = -4.1$ for the
differential size distribution between at least 3 AU before
and 1.5 AU after perihelion in 2002 \cite{agarwal-mueller2007}.
%
%
To put this result in a wider context, we have compared 
the measured brightness distribution along the antitails of different comets rather than 
the derived size distributions, because we found that different authors use
different models to derive the size distribution from the measured brightness
exponent.
%
We found that the brightness variation measured for the antitail of 67P/C--G
is rather typical compared with 
the antitails of other comets, such that also the dust size distribution of
67P/C--G may be typical.


\subsection{Trail Analyses}
\label{subsec:model_trail}
Several wide field images of the dust along the orbit of \cg\ were obtained
between 2002 and 2006 in both optical and infrared light, and their analyses
are summarised in the following. 

Three images -- one optical and two infrared at 24\,\micron\ -- with fields of view of about half a degree were analysed using a
generalised Finson-Probstein approach that takes into account
the tidal deformation of dust shells on long time-scales \cite{agarwalPhD}. 
The optical image was
obtained in April 2004 with the Wide Field Imager at the ESO/MPG 2.2m
telescope on La Silla at \rh\ = 4.7 AU (Figure~\ref{fig:trail}), and the infrared images were taken by the MIPS
instrument on board the Spitzer Space Telescope of NASA in August 2005 and
April 2006 (\rh = 5.7 AU in both cases).  All three images were thus taken
when the comet was close to aphelion and not active, such that no particles
larger than approximately 100\,\micron\ are expected to be present in the fields of view.
Simulated images were generated taking into account 
dust emitted during all seven perihelion passages since the last close
encounter with Jupiter in 1959. The time-dependence of
the dust production was modelled on the observed time-dependence of \afr.
The
emission speeds and surface activity were assumed to be isotropic, and the
relative dependence of the emission speeds on size and heliocentric
distance was obtained by help of the hydrodynamic coma model described in
Section~\ref{sec:mmmodel} of this article, and from the observed
time-dependence of the water production rate. 
Within this framework, the parameter values found most suitable to reproduce
both the brightness profile along the trail and its width 
were the following: The emission speeds of a 1\,mm radius particle at
perihelion ranged from 6\,m/s to 12\,m/s, decreasing by a factor of 10 until 3
AU. The size dependence of $\beta$ could be characterised by Equation~\ref{eq:beta}
with $Q_{\rm pr}/\rho$ between 1 and 3 cm$^3$/g, which -- assuming $Q_{\rm
  pr}$ = 1 for large particles -- corresponds to bulk densities between
0.3 and 0.9 g/cm$^3$. The size distribution exponent for particles larger than
100\,\micron\ was in the range of -4.1 to -3.9. In order to reproduce
both the optical and the infrared images, a geometric albedo for visible light
of $p$ = 4\% was required if the dust was assumed to be emitting as a blackbody
($\epsilon$ = 1 in Equation~\ref{eq:T_eq}), but $p$ = 10\% if $\epsilon / (1-A)$ =
0.6 \cite{hanner-tedesco1985a, sykes-walker1992a}. 
Lower limits for the production rates of particles with $s$ $>$ 100\,\micron\
ranged from 100 kg/s at perihelion to
0.2 kg/s at 3 AU, corresponding to \afr\ values of 4 and 0.05\,m,
respectively. This implies that particles larger than 100\,\micron\ would have
contributed at least 50\% of the total \afr\ observed while the comet was in
the inner solar system, which is difficult to reconcile with the size
distribution exponent on the order of -4.

Three optical images (R-band) were obtained between September
2002 and February 2003 with the 1.05-m Schmidt telescope of the Kiso
Observatory at 
Nagano, Japan \cite{ishiguro2008}. The difference in position angle of freshly
emitted dust and the trail (dust from previous apparitions) was 1\degree\ or
larger in these images, thus -- in contrast to images taken at later dates --
a clear separation of both dust populations is possible. Taking into account
dust emitted after aphelion in 1986 and 
assuming $p$ = 4\%, $\rho$ = 1 g/cm$^3$,
and $Q_{\rm pr}$ = 1, the images were best reproduced by a model
with cone-shaped emission with a half opening angle between 45\degree\ and
90\degree, a size distribution exponent of -3.5 with particles in the range
between 6\,\micron\ and 5\,mm, and dust production rates of about 200 kg/s at
perihelion and 15 kg/s at 3 AU. The emission speeds were assumed to be proportional
to $s^{-1/2}$ and $r_{\rm h}^{-1/2}$, absolute values for a 1\,mm particle
ranging from 8 to 18 m/s at perihelion and from 5 to 12 m/s at 3 AU. 

Two further observations were analysed by help of a Monte Carlo model \cite{kelley-reach2008}:
an optical image (Gunn r' filter) obtained in June 2003 with the 5-m Hale telescope at
Palomar Observatory, and an infrared (24\,\micron) image made with Spitzer/MIPS
in February 2004. The trail of particles from previous
perihelion passages was not detected in the Palomar image, but was visible to
Spitzer.
The applied model includes the assumptions of $Q_{\rm pr}$ = 1,
$\rho$ = 1 g/cm$^3$, dust production rates $Q_{\rm d} \propto r_{\rm
  h}^{-5.8}$, and emission speeds $v \propto \sqrt{\beta/r_{\rm
    h}}$. Particles larger than 0.5\,\micron\ and emitted after March 1993
were included in the analysis. The images were best fitted by a cosine-shaped
distribution of the surface activity, peaking at the subsolar point, and a
differential size distribution exponent of -3.5. The emission speeds of a 1\,mm
particle at the subsolar point varied between 10 m/s at perihelion and 7 m/s
at 3 AU.

\subsection{Summary of Modelling Results}
\label{subsec:model_summary}
\begin{figure}[h]
\includegraphics[width=.5\textwidth]{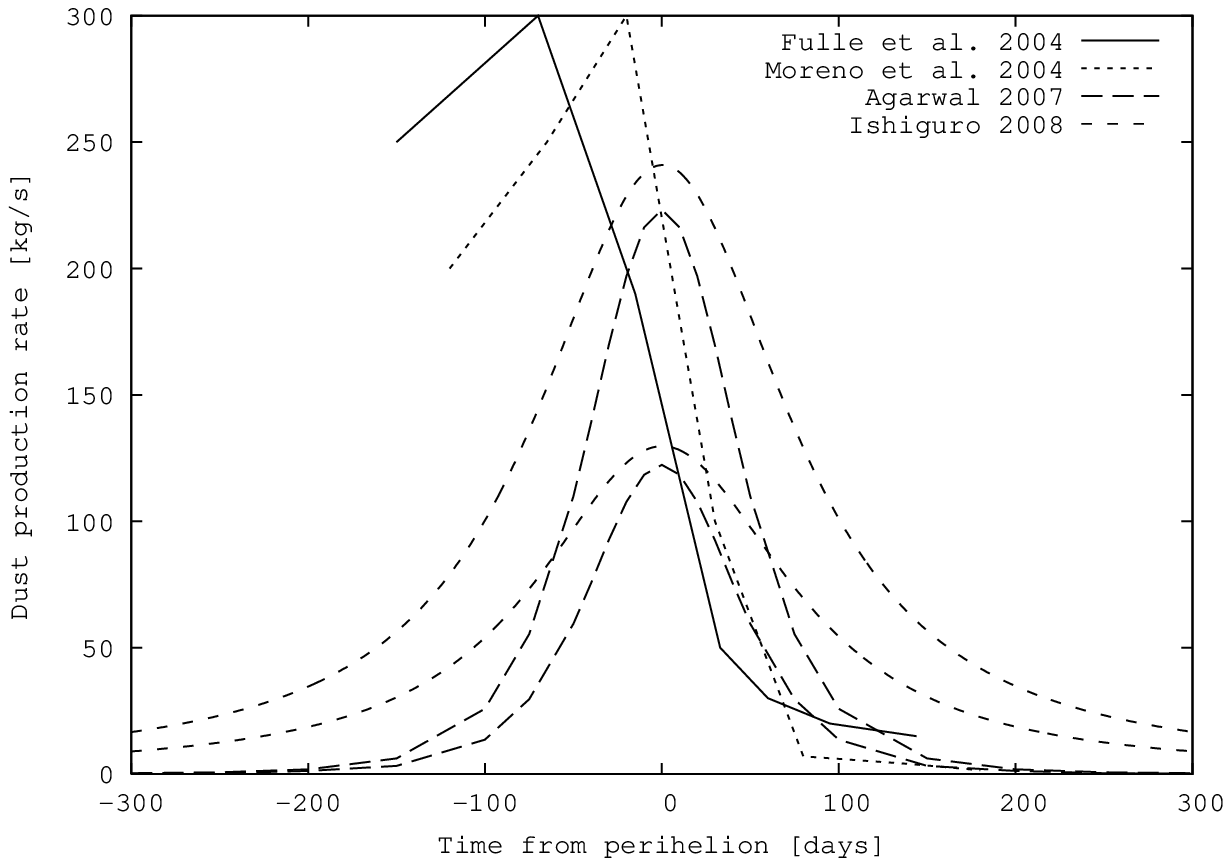}
%
\includegraphics[width=.5\textwidth]{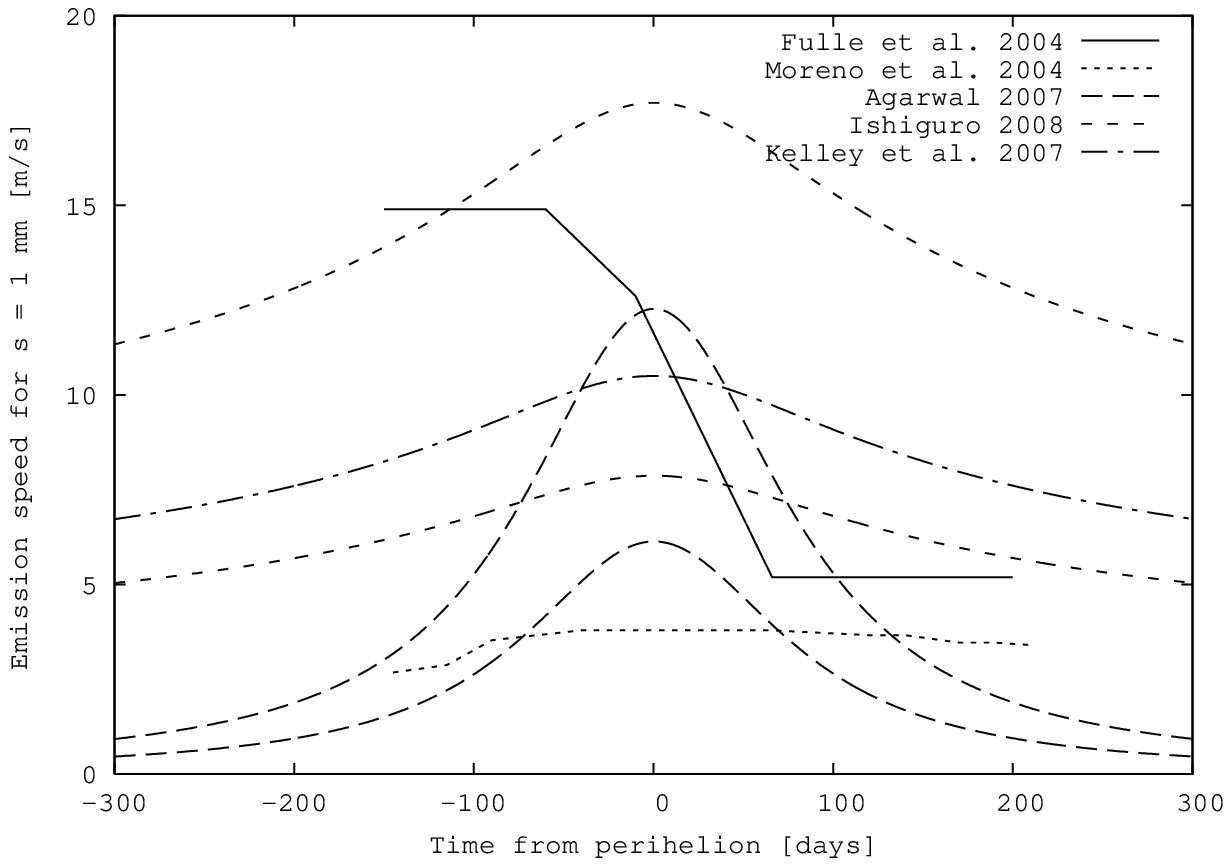}
%
\begin{center}
\includegraphics[width=.5\textwidth]{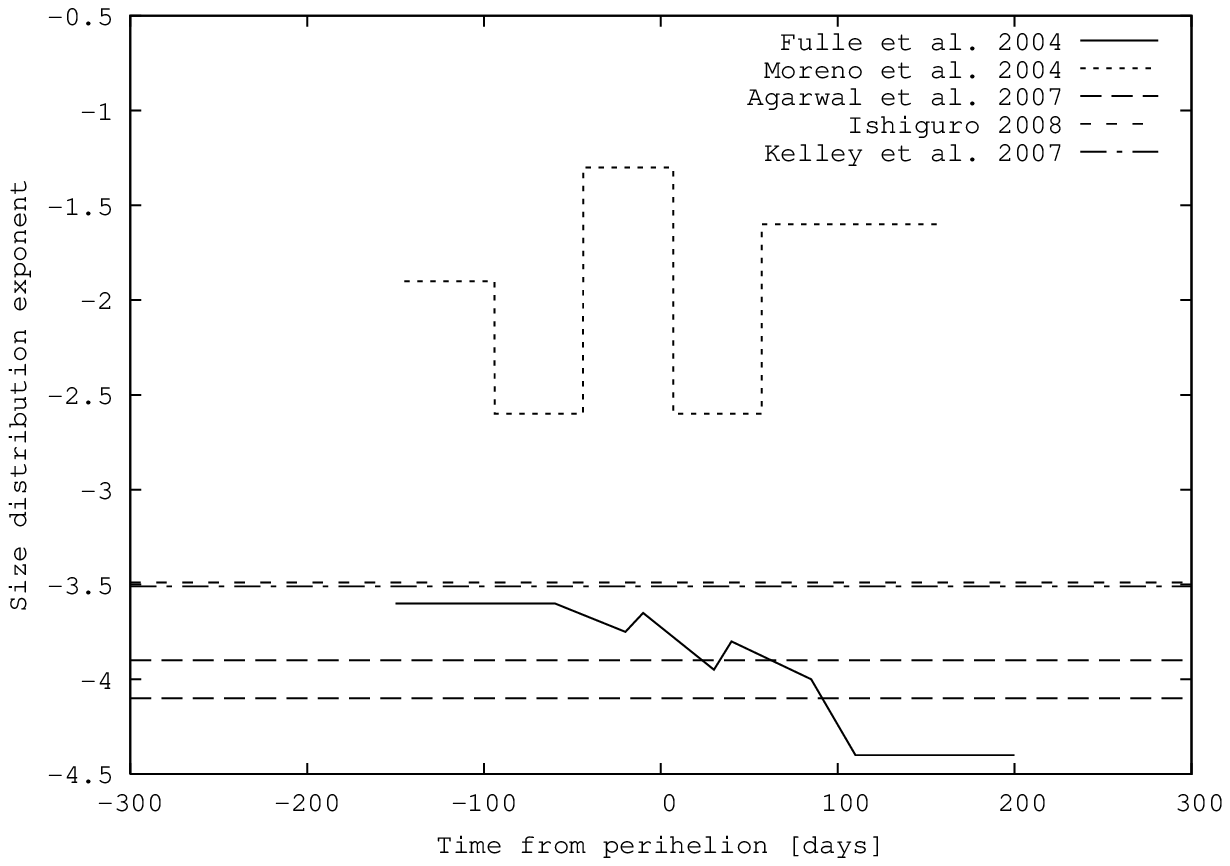}
\end{center}
\caption{Comparison of dust characteristics derived from the different models
  described in Section~\ref{sec:model}. Two lines of the same style represent
  upper and lower limits. Top left: dust production rates. Top right: emission
  speeds of a dust grain with 1\,mm radius. For anisotropic
  models, the peak emission speed (assumed above the subsolar point) is
  given. Bottom: Exponents of the differential size distribution.}
\label{fig:comparison_model}
\end{figure}

Figure~\ref{fig:comparison_model}
summarises the dust production rates, emission speeds and size distributions
derived by help of the various models described in
Sections~\ref{subsec:model_fulle} to \ref{subsec:model_trail}. We wish to stress that -- with the exception of the
TNG image obtained in March 2003 -- each model has been applied to a different
set of images, and each model was able to reproduce the images that were
analysed by it. However, the ranges of parameter values derived from the
different models are considerable, such that it is at present difficult to derive a
consistent picture of the CG dust environment on the basis of these
results. Future work on this matter should focus on finding a set of
parameters that is able to reproduce all available observations of the CG
tail and trail.

\section{Dust Environment Model and Predictions for 67P/C--G}
\label{sec:mmmodel}
In this section, the state of dust in the coma of \cg\ is predicted by means
of a hydrodynamic model
\cite{mueller1997,mueller1998,muellerPhD,landgraf-mueller1999}.
Three emission scenarios and three size distribution exponents are studied,
and estimates of the fluxes on a spacecraft
for several types of trajectory and the radiation environment in the coma
are presented.

\subsection{Coma Model and Parameters}
\label{subsec:mmmodel_param}
The employed model can be
considered as the simplest possible physically consistent model of the inner
coma: The comet nucleus is assumed to be of spherical shape, and the activity
distribution on its surface is axis-symmetric with respect to the comet-Sun
line. The flows of gas and dust around the comet are
calculated rendering at any
position in the coma the gas density, velocity, and temperature, and
the number density and velocity of dust particles of different mass
classes. Each class represents particles having masses within one decade in
the range of $10^{-20}$ to $10^{4} \,{\rm kg}$.
All dust particles of one class are assigned the same representative
mass and radius. It is assumed that the particles have spherical shapes and a
constant bulk density of $1000 \, {\rm kg/m}^3$.

The numerical method used to compute the dynamics of gas and dust takes
advantage of the fact that the influence of the dust on the gas flow is of
minor importance. In a first step, the gas flow is calculated
without taking into account the presence of dust. In a second step, dust
trajectories are integrated considering the gravitation of the nucleus and
the gas drag force. 

The gas flow is calculated under the assumptions that 
the gas is in thermal equilibrium everywhere, and that the mean molecular mass is constant
across the coma. It is given by the mean of the masses of the two
most common molecules, H$_2$O and CO, weighted by their overall
abundances in the coma. 
The production rates of these species are input parameters to the
model. To first order, the CO-activity can be treated as independent
of heliocentric distance, and the constant value of $10^{26}$ molecules/s is
adopted for 67P/C--G. An upper limit for the CO-production of $10^{27}$ molecules/s at
3 AU has been inferred from radio observations \cite{bockelee-moreno2004}. The water activity is assumed
to vary with heliocentric distance and time as indicated by observations
(Figure~\ref{fig:afr_h2o}, bottom). 
The dust activity is scaled such that the observed \afr\ is reproduced for the
employed size distribution and albedo. 

\begin{figure}[hpt]
\includegraphics[width=.5\textwidth]{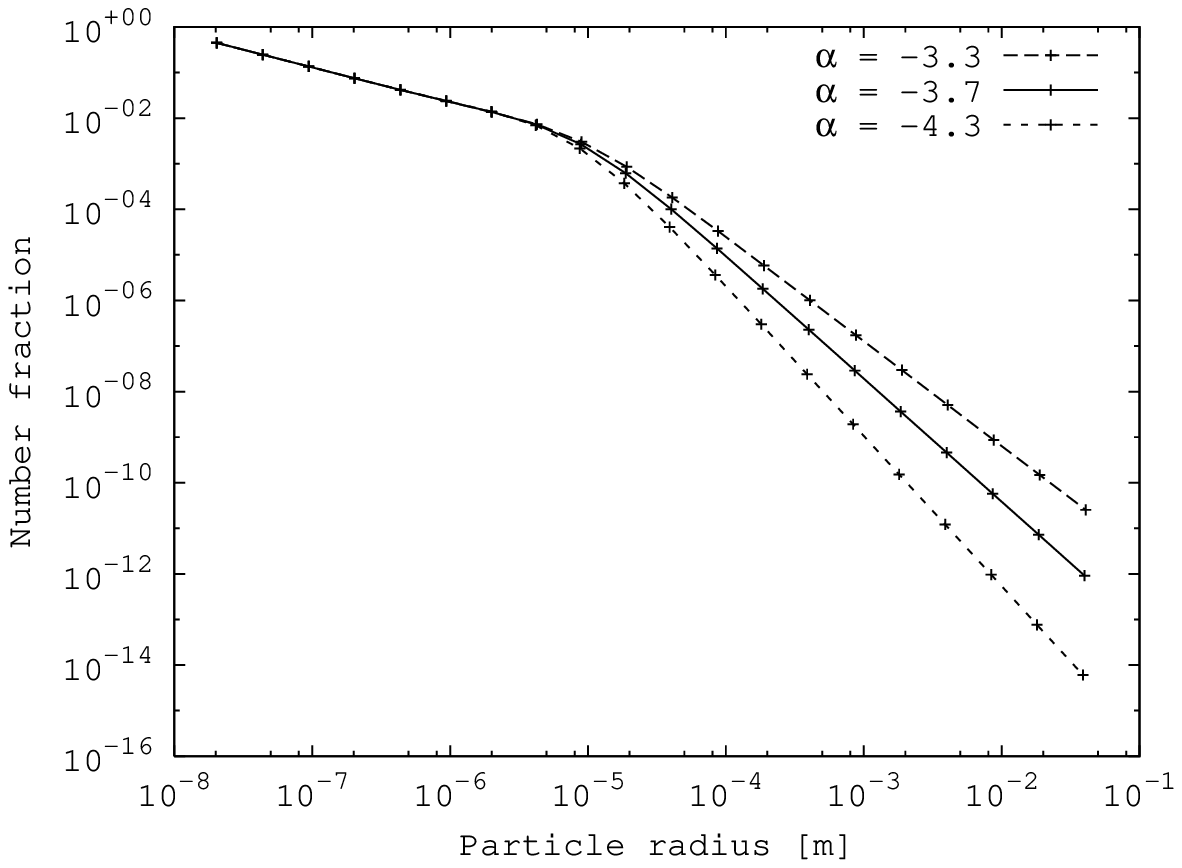}
%
\includegraphics[width=.5\textwidth]{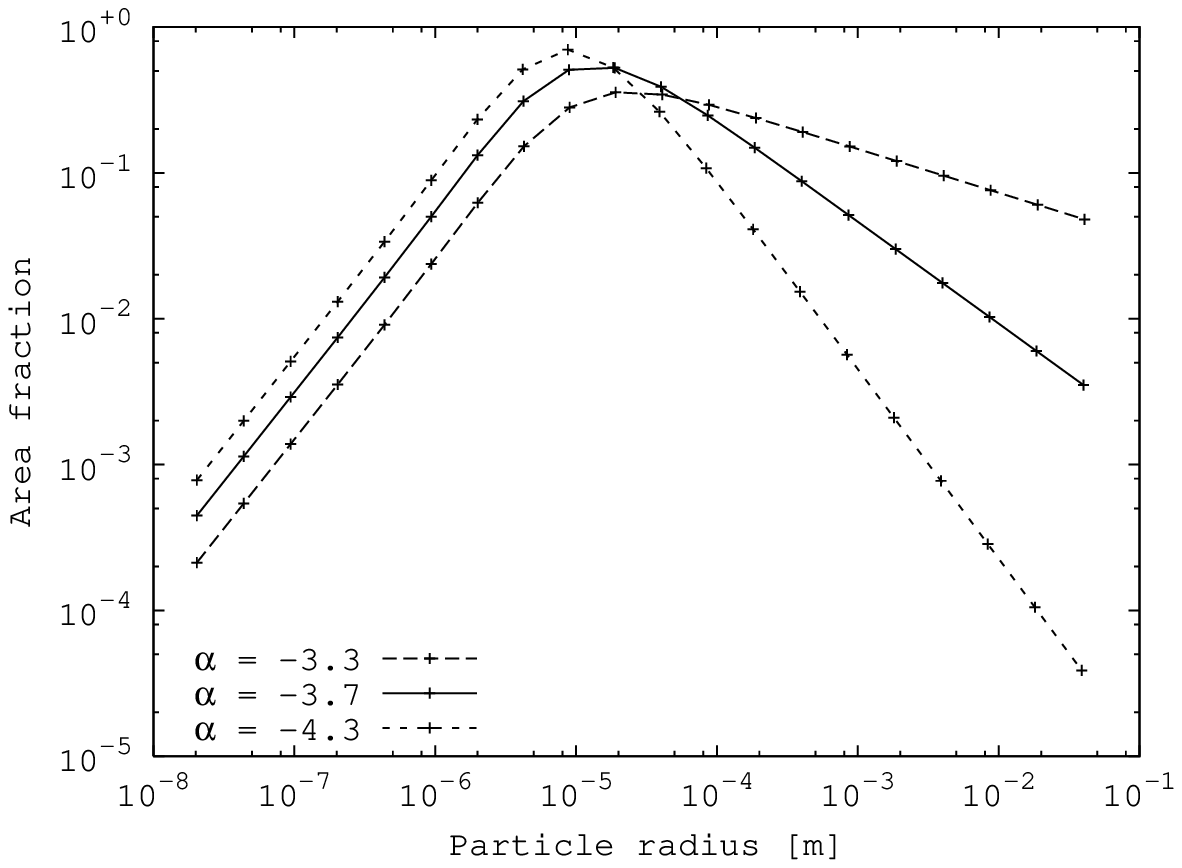}
%
\begin{center}
\includegraphics[width=.5\textwidth]{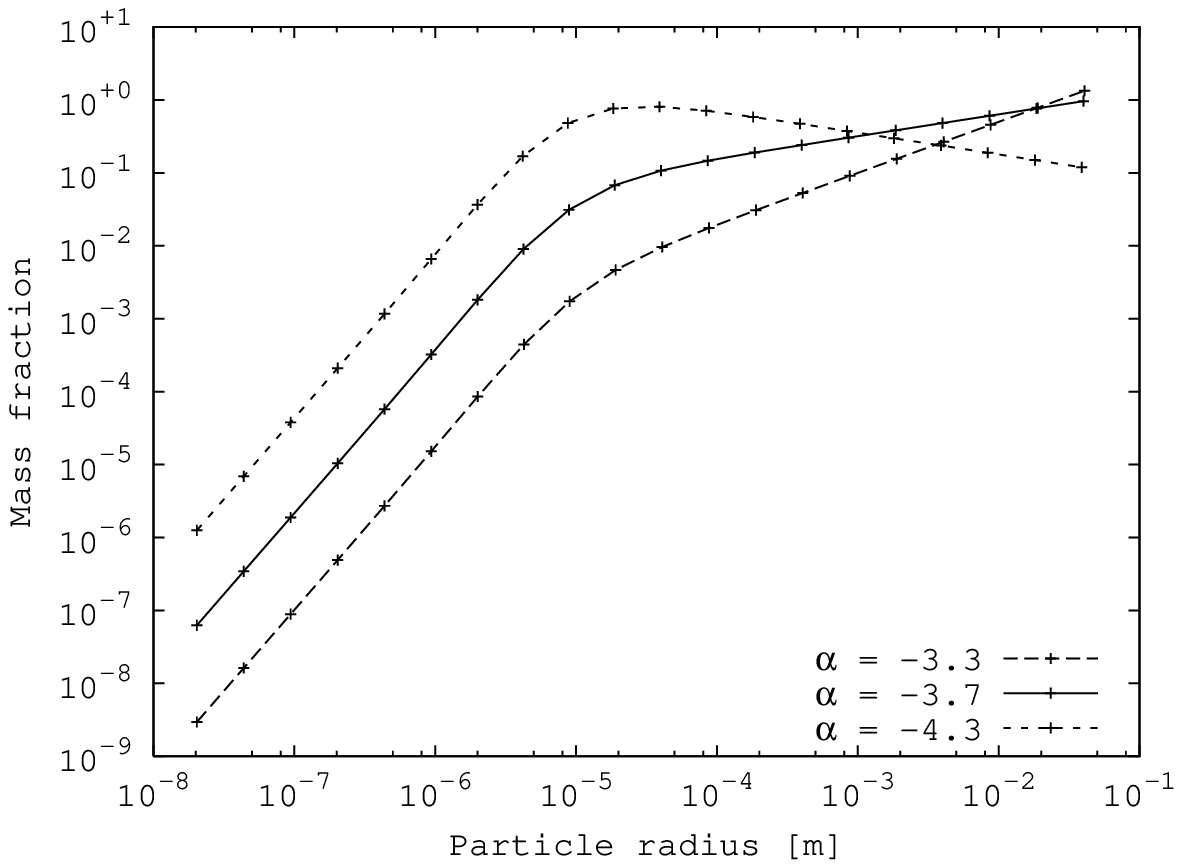}
\end{center}
\caption{Different moments of the mass distributions studied in
  Section~\ref{sec:mmmodel}. Top left: number fraction. Top right: area
  fraction. Bottom: mass fraction. To convert mass to radius, a bulk density
  of $\rho$ = 1000 kg/m$^3$ was assumed.}
\label{fig:md_moments}
\end{figure}

Since, at present, the size distribution estimates vary considerably
(Section~\ref{subsec:model_summary}), we consider in the following
the same range of mass distributions as in earlier
models for 46P/Wirtanen \cite{mueller1997}: two
``extreme'' and a ``nominal'' one. 
All are characterised by different exponents for light and heavy
particles, and the analytical form used for the cumulative mass distribution
is as follows \cite{divine-newburn1987b}:
\begin{equation}
\label{eq:divnew_md}
F(m) = \left( \frac{(1+x)^{b-1}}{x^b} \right)^{ac} 
{\rm with} \:\: x = \left( \frac{m}{m_{\rm t}} \right)^{1/c}.
\end{equation}
This function is specified by the positive parameters $a$, $b$, $c$, and
$m_{\rm t}$. 
The exponent $-\gamma$ of the cumulative mass distribution tends towards
$-a$ for heavy particles ($m \ll m_{\rm t}$) and towards $-ab$ for light
ones. $m_{\rm t}$ is the mass where the transition between the two exponents
takes place, and $c$ determines the sharpness of the transition.
The three mass distributions considered in this paper are different only by
their slopes for heavy particles. For light particles we use $\gamma = ab =
0.26$ throughout (corresponding to $\alpha = -1.8$), for the transition mass
$m_{\rm t} = 10^{-13} \,{\rm kg}$, and $c = 2$. The slopes for large
particles are as follows: the (velocity-corrected) fit to the fluence
measured on Giotto \cite{divine-newburn1987b} gave $\alpha = -4.3$ ($\gamma =
1.1$) which we use as one of the ``extreme'' distributions. From the same data, an exponent in the range of $\alpha \in [-3.7,
  3.3]$ has been derived \cite{fulle-colangeli1995}, wherefore we adopt
$\alpha = -3.7$ ($\gamma = 0.9$) as the ``nominal'' and $\alpha = -3.3$
($\gamma = 0.8$) as the other ``extreme'' distribution. It must, however, be
emphasised that for $\alpha > -3.5$,
the observed brightness is dominated by light scattered by large particles
\cite{mueller1997}. 
Figure~\ref{fig:md_moments}  show
different moments of the mass distribution.

In the following, model predictions are given for several heliocentric
distances: the perihelion distances in 2015 ($r_{\rm h} = 1.24 \, {\rm
  AU}$) and 2002 ($r_{\rm h} = 1.29 \, {\rm AU}$), and   
 $1.87 \, {\rm AU}$, $2.5 \, {\rm AU}$, and $3 \, {\rm AU}$. The values
used for the water production rate, for \afr, and for the phase angle $\alpha$
are listed in Table~\ref{tab:prodrates}. We use a geometric albedo of 4\% and the phase function shown in Figure~\ref{fig:phase_fns}.
\begin{table}[b]
\caption{Water production rates ${\rm Q}_{{\rm H}_2{\rm O}}$, \afr-values and
  phase angles $\alpha$ for those
  heliocentric distances $r_{\rm h}$ used to obtain
  the model predictions in this section.}
\centering
\label{tab:prodrates}
\begin{tabular}{llll}
\hline\noalign{\smallskip}
$r_{\rm h}$ [AU] & ${\rm Q}_{{\rm H}_2{\rm O}}$ [molecules/s] & $Af\!\rho \,
(\alpha) \, [{\rm cm}]$ & $\alpha [{\rm deg}]$ \\[3pt]
$1.24$ & $1.3 \times 10^{28}$ & $400$ & 35 \\
$1.29$ & $1.0 \times 10^{28}$ & $300$ & 35 \\
$1.87$ & $1.2 \times 10^{27}$ & $65$ & 32 \\
$2.5$ & $1.8 \times 10^{26}$ & $20$ & 5 \\
$3.0$ & $6.4 \times 10^{25}$ & $5.8$ & 19 \\
\noalign{\smallskip}\hline

\end{tabular}
\end{table}
While  
the production of CO is assumed to be isotropic, 
for the release of H$_2$O three different surface activity distributions are
taken into account. The
first represents a homogeneously composed surface such that the strength of gas
production is determined by the amount of solar energy received
by a given surface element ({\em insolation driven model}\,). In the second
scenario, a {\em jet} is simulated describing the surface activity by a
Gaussian profile centred at the subsolar point
\cite{fulle-colangeli1995,agarwal-mueller2004}. This model is expected to yield upper limits for
local quantities inside a jet, while the insolation driven model gives
estimates for quantities averaged over time and space. In addition, an even
simpler {\em radially symmetric model} is used in which the water
activity is distributed equally over the comet surface. 

The average radius of the nucleus of 67P/C--G as given in the literature is in
the range of $1.7 \, .. \, 3.2 \,{\rm km}$
\cite{mueller1992,lowry-fitzsimmons2003,tancredi-fernandez2000,tancredi-fernandez2006,lamy-toth2007,tubiana-drahus2007_dps,kelley-reach2008}.
The geometric albedo of the nucleus in R-band is in the range of $0.045$ to
$0.06$, and
the bulk density of the nucleus is estimated to $370 \,{\rm kg/m^3}$ \cite{lamy-toth2007}. 
The following results were obtained for a spherical nucleus
having a radius of 2 km and a geometric albedo of 4\%. 

\subsection{Dust Terminal Speeds}
\label{subsec:mmmodel_speeds}
\begin{figure}[t]
\centering
\includegraphics[width=.7\textwidth]{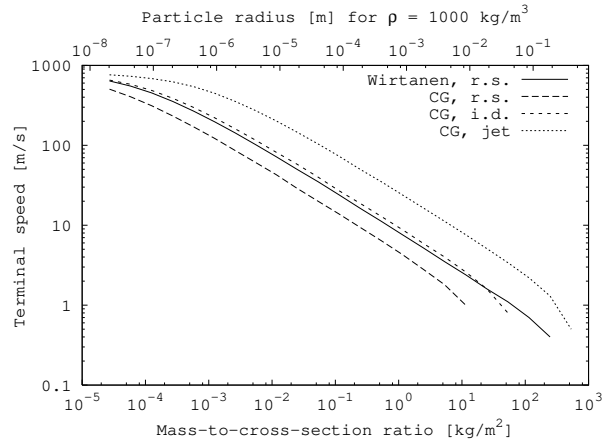}
\caption{ 
Terminal speeds as functions of mass-to-cross-section ratio for various emission scenarios (r.s.: radially
symmetric, i.d.: insolation driven). The corresponding particle radii for a
bulk density of $\rho = 1000 \, {\rm kg/m^3}$ are indicated at the upper
margin. 
Values for 67P/C--G are
given at 20 km from the nucleus centre above the subsolar point and at perihelion ($r_{\rm h} =
1.29\,{\rm AU}$). For 46P/Wirtanen they are at 10 km from the nucleus centre
and at the perihelion distance of $r_{\rm h} = 1.06\,{\rm AU}$.}
\label{fig:termvel}
\end{figure}

The terminal speeds of dust as a function of particle size for the three
different emission scenarios are shown in Figure~\ref{fig:termvel}. For
comparison, the corresponding values for comet 46P/Wirtanen, the former
Rosetta target, are shown as well. Note that all curves are at
perihelion, hence the comets are compared at different
heliocentric distances. While the terminal speeds for 46P/Wirtanen are given at 10 km
from the nucleus centre, those for 67P/C--G are at 20 km, because the dust
particles have not quite reached their terminal speeds at lower heights. 

In the radially symmetric models, the terminal speeds
are higher in the case of 46P/Wirtanen than of 67P/C--G. The reason is
that both comets have similar gas production rates although
46P/Wirtanen is smaller. Hence the gas directly over the surface is denser and
the dust reaches higher velocities. The ratio between the speeds expected for
67P/C--G and 46P/Wirtanen in the radially symmetric model ranges from 0.8 for
small particles to 0.5 for large ones. 
%
%
The speeds shown in Figure~\ref{fig:termvel} were computed for a dust bulk
density of $\rho = 1000 \,{\rm kg/m}^3$. Scaling to other particle densities
can to first order be done assuming that particles with the same cross section
to mass ratio reach the same velocities. 
For the larger particles, a power law of the following form was fitted to the
calculated speeds:
\begin{equation}
\label{eq:termvel_fit}
v\,(s) = v_0 \left(\! \frac{s}{s_0} \!\right)^{\!-d},
\end{equation}
with $s_0 = 1 {\rm mm}$.
The resulting parameters for the different scenarios are given in
Table~\ref{tab:termvel_fit}. For the fits, all dust classes with particle
radii $s > 0.1 \, {\rm mm}$ have been used. The class with the largest
liftable particles was not included because of the strong influence of the
gravitation of the comet. 
For 46P/Wirtanen, the exponent $d$ is closer to the expected value of 0.5
than for 67P/C--G because of the latter's larger nucleus and
resulting stronger gravity.  

\begin{table}[h]
\caption{Results of fitting a power law to the dust terminal speeds for large
 particles above the subsolar point at perihelion. Values are given for 
different emission models: radially symmetric (r.s.), insolation driven (i.d.), and jet model.}
\centering
\label{tab:termvel_fit}
\begin{tabular}{llll}
\hline\noalign{\smallskip}
Model & $r_{\rm h}$ [AU] & $v_0$ [m/s] & Exponent $d$ \\[3pt]
Wirtanen r.s. & 1.06 & 7.0 & 0.51 \\
C--G r.s. & 1.29 & 3.9 & 0.53 \\
C--G i.d. & 1.29 & 8.0 & 0.53 \\

C--G jet & 1.29 & 22.2 & 0.52 \\
\noalign{\smallskip}\hline

\end{tabular}
\end{table}

\subsection{Dust Flux at 1.29 AU}
\label{subsec:mmmodel_flux}
\begin{figure}[h]
\includegraphics[width=.5\textwidth]{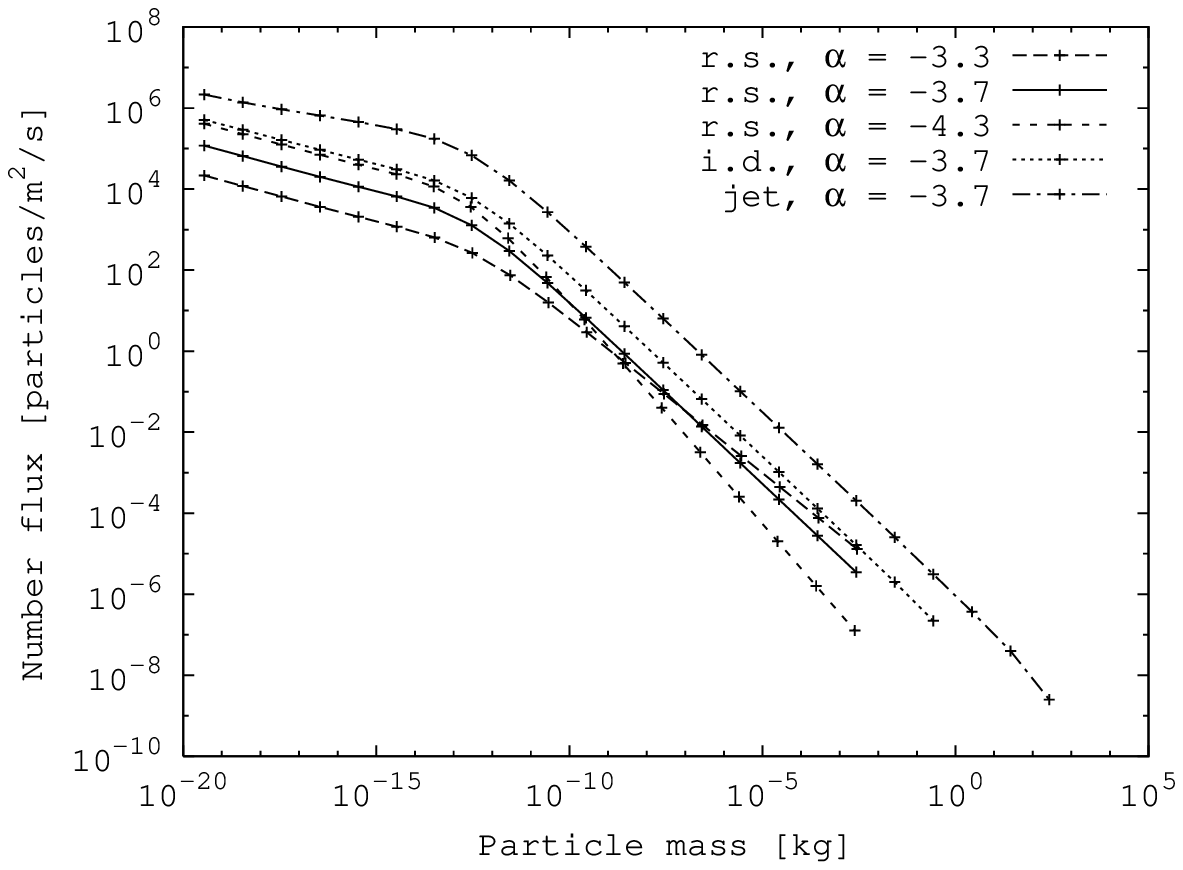}
%
\includegraphics[width=.5\textwidth]{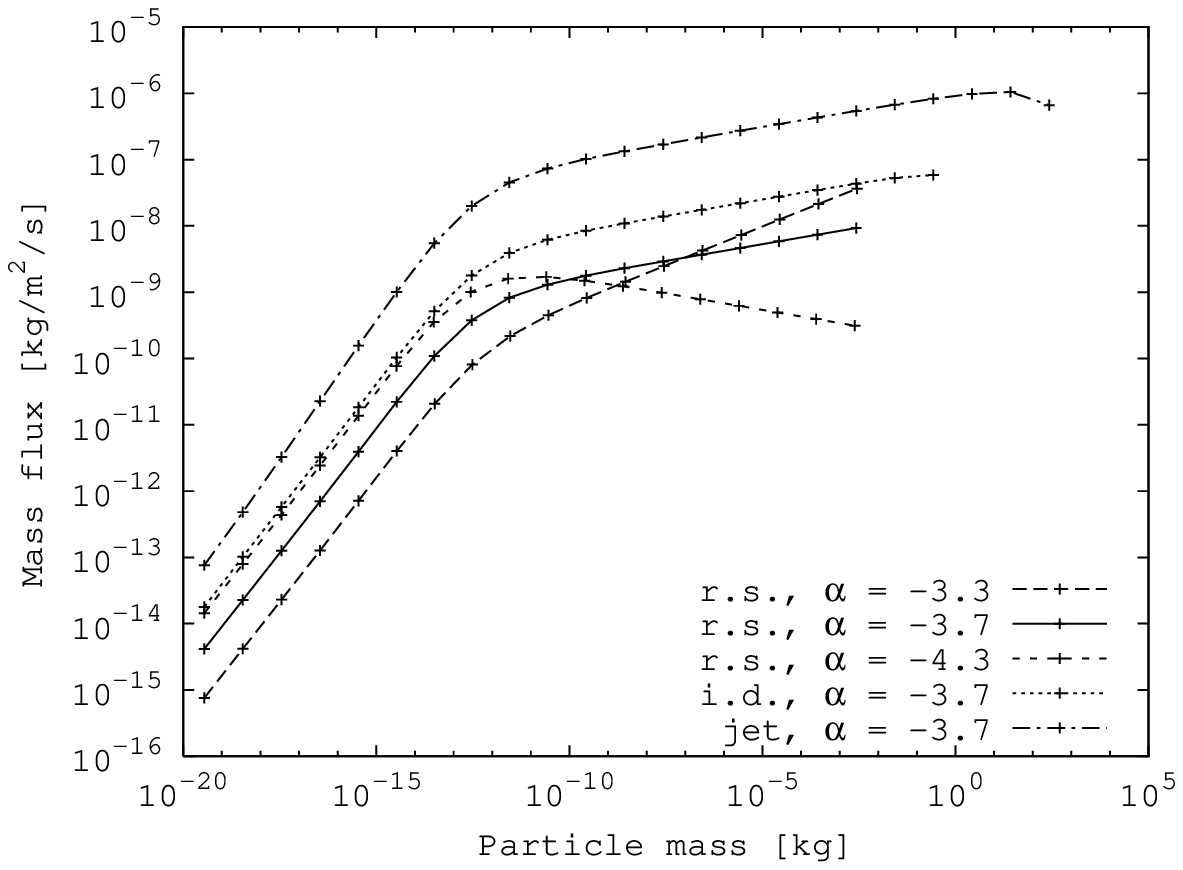}
%
\includegraphics[width=.5\textwidth]{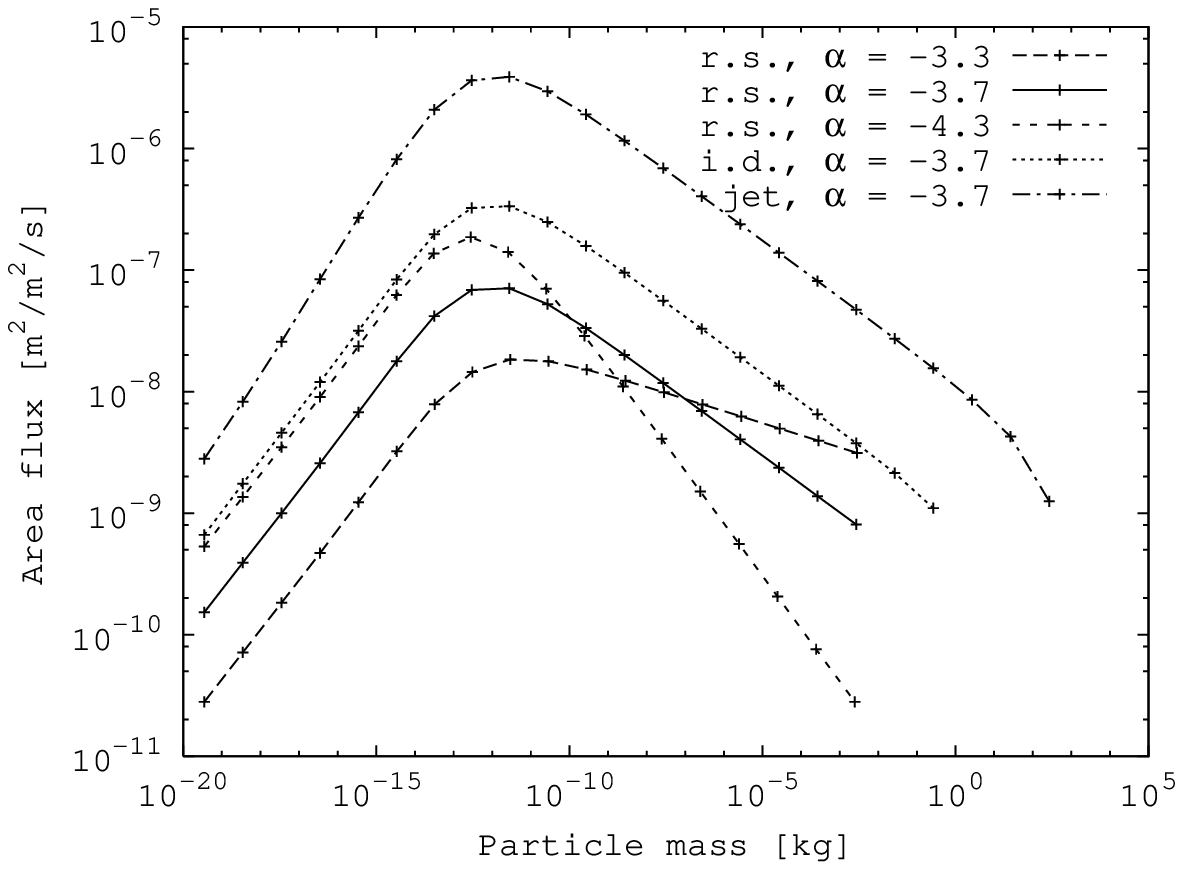}
%
\includegraphics[width=.5\textwidth]{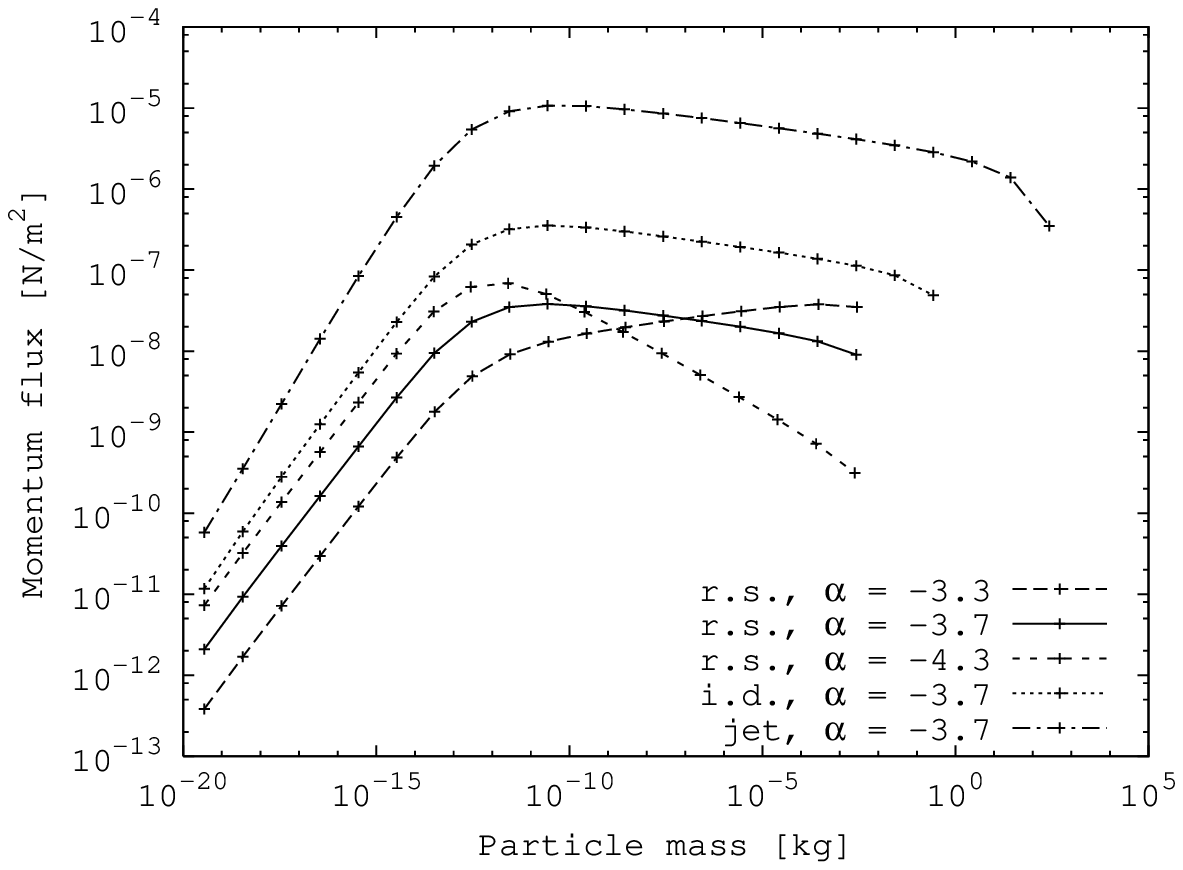}
\caption{Fluxes over particle mass for various mass distributions and emission
  geometries. Top left: number flux, top right: mass flux, bottom left: area
  flux, bottom right: momentum flux.}
\label{fig:fluxes}
\end{figure}

Here we give numerical values for the fluxes of gas and dust on a surface
directed towards the nucleus at 20 km from
the nucleus centre above the subsolar point for various mass distributions and
emission scenarios. As the speeds of gas and dust outside the
acceleration zone are almost constant, the fluxes there can be scaled by an
inverse square law.
Note that the values given for the gas are only valid as long as the gas is
cold, i.e. far from the nucleus. Within a few comet radii from the surface, the
velocity distribution of the gas 
should be taken into account. 
Consequently, there will also be a flux on surfaces which are directed away
from the flux direction. The numerical values are given in
the appendix of \cite{agarwal-mueller2007} and visualised
in Figure~\ref{fig:fluxes}. 
The number and mass fluxes (upper panels Figure~\ref{fig:fluxes}) in are relevant for different sensors of the GIADA
instrument. The lower left panel shows the surface coverage by dust
which is important for estimating thermal effects on spacecraft and
instruments. The lower right panel describes the momentum transfer onto
the spacecraft by the dust. 
 
\subsection{Dust Collection along Sample Spacecraft Trajectories}
\label{subsec:mmmodel_traj}
During most of the mission, regions of strong dust emission in the coma 
will be avoided, because impacts of dust particles constitute a hazard 
to the Rosetta spacecraft. This hazard includes damage of sensitive 
surfaces by cratering and erosion due to high velocity dust impacts, 
contamination of surfaces 
and modification of their optical and thermal properties due to sticking of
dust, and even mechanical blockage of gears  
and hinges by larger grains. 
This hazard has been taken into account in the design of the Rosetta
spacecraft and its instruments, e.g. by painting the spacecraft black so that
sticking dust will not significantly change the thermal
properties of the spacecraft, and by employing movable shutters in front of
sensitive surfaces such as cameras. Nevertheless, in the vicinity of the
comet, spacecraft operations will
keep exposure to dust to a minimum \cite{koschny-dhiri2007}. In order to fulfil the
dust collection requirements of the COSIMA and MIDAS instruments the
spacecraft  
occasionally will have to pass through regions of high dust 
density. The design of these passages will depend on the 
collection requirements of the dust instruments.

For simple dust collection trajectories, the 
surface coverage by dust (assuming all dust sticks to the surface) has 
been calculated for the different dust emission scenarios and mass 
distributions described in Section~\ref{subsec:mmmodel_param}. The considered
trajectory types are one 
revolution on a circular orbit, a parabolic orbit, and a straight line 
trajectory passing the nucleus at a speed v = 1 m/s. All 
trajectories are assumed to pass at closest approach ($r_{\rm
  CA}=4$\,km from the nucleus centre)
over the subsolar point, i.e. the region of 
highest dust density in our model. The fraction of the surface covered by dust after 
one passage of the nucleus is given in Table~\ref{tab:sc_fluences}. 
\begin{table}[b]
\caption{Total surface coverage of the nucleus-directed side of the spacecraft due to dust particles of all sizes.
  The table lists the surface coverage for different heliocentric distances
  and different kinds of orbits.
At heliocentric distances of 2.5 and 3~AU the water production rate is
  comparable with the background CO production rate, therefore, dust
  emission in form of a jet is not considered.
  We expect that at these heliocentric distances particles on the order of
  100\,\micron\ can still leave the comet.
  All orbits have the point of closest approach to the nucleus over the subsolar point with 
  a distance $r_{\rm CA}=4$\,km from the nucleus centre. The considered types of orbits are a parabola,
  a straight line at constant speed of 1 m/s, and a circle. For the parabola and the circle only the part of
  the trajectory over the comet day side is taken into account. }
\centering
\label{tab:sc_fluences}
\begin{tabular}{@{\extracolsep{1mm}}llllllll}
\hline\noalign{\smallskip}
Model & Orbit &  Size &&& $r_{\rm h}$ && \\
&&               distr.&&& [AU] &&  \\
&&               exp.  & 1.24 & 1.29 & 1.87 & 2.5 & 3.0 \\[3pt]
radial & circle   & -3.3 & 0.101 & 0.075 & 0.006 & 0.0008 & 0.0004 \\
sym.   &          & -3.7 & 0.272 & 0.204 & 0.015 & 0.0015 & 0.0005 \\
       &          & -4.3 & 0.539 & 0.404 & 0.030 & 0.0028 & 0.0009 \\
insol. & parabola & -3.3 & 0.189 & 0.142 & 0.017 & 0.0016 & 0.0005 \\
driven &          & -3.7 & 0.606 & 0.455 & 0.035 & 0.0026 & 0.0007 \\
       &          & -4.3 & 1.213 & 0.909 & 0.062 & 0.0042 & 0.0010 \\
       & straight & -3.3 & 0.148 & 0.111 & 0.013 & 0.0013 & 0.0004 \\
       & line     & -3.7 & 0.473 & 0.355 & 0.027 & 0.0021 & 0.0006 \\
       &          & -4.3 & 0.945 & 0.709 & 0.048 & 0.0034 & 0.0009 \\
jet    & parabola & -3.3 & 0.917 & 0.688 & 0.106 &        &        \\
       &          & -3.7 & 3.540 & 2.655 & 0.249 &        &        \\
       &          & -4.3 & 6.896 & 5.172 & 0.421 &        &        \\
       & straight & -3.3 & 0.685 & 0.514 & 0.079 &        &        \\
       & line     & -3.7 & 2.644 & 1.983 & 0.185 &        &        \\
       &          & -4.3 & 5.153 & 3.865 & 0.313 &        &        \\
\noalign{\smallskip}\hline
\end{tabular}
\end{table}
For other closest 
approach distances the surface coverage levels scale with $r_{\rm CA}^{-0.5}$ for 
Keplerian orbits (circles and parabola) and with $r_{\rm CA}^{-1}$ and $v^{-1}$ for 
straight line passages. The values in Table~\ref{tab:sc_fluences} refer to different dust 
production rates, i.e. different heliocentric distances. 

A surface coverage of one implies roughly a mono-layer of differently
sized dust grains, most of which are grains in the 1 to 10 \micron\ size 
range (cf. Figure~\ref{fig:md_moments}, top right). Since the dust instruments are sensitive to 
different sizes of dust grains it is useful to determine the dust 
coverage by particles above a certain minimum size. This is achieved by
applying the scaling factors for different minimum dust sizes given in Table~\ref{tab:areacum}. 
\begin{table}[t]
\caption{Fraction of area flux due to particles greater than mass $m_{\rm
    min}$ or radius $s_{\rm min}$ for the different mass distributions.}
\centering
\label{tab:areacum}
\begin{tabular}{@{\extracolsep{2mm}}lllll}
\hline\noalign{\smallskip}
$m_{\rm min}$ [kg]& $s_{\rm min}$ [m]& $\alpha = -3.3$ & $\alpha=-3.7$ & $\alpha=-4.3$ \\[3pt]
 $1.000\times 10^{-20}$ & $1.337\times 10^{-08}$ & $1.000\times 10^{+00}$ & $1.000\times 10^{+00}$ & $1.000\times 10^{+00}$ \\
 $1.000\times 10^{-19}$ & $2.879\times 10^{-08}$ & $9.998\times 10^{-01}$ & $9.996\times 10^{-01}$ & $9.992\times 10^{-01}$ \\
 $1.000\times 10^{-18}$ & $6.204\times 10^{-08}$ & $9.993\times 10^{-01}$ & $9.984\times 10^{-01}$ & $9.972\times 10^{-01}$ \\
 $1.000\times 10^{-17}$ & $1.337\times 10^{-07}$ & $9.979\times 10^{-01}$ & $9.955\times 10^{-01}$ & $9.921\times 10^{-01}$ \\
 $1.000\times 10^{-16}$ & $2.879\times 10^{-07}$ & $9.945\times 10^{-01}$ & $9.880\times 10^{-01}$ & $9.788\times 10^{-01}$ \\
 $1.000\times 10^{-15}$ & $6.204\times 10^{-07}$ & $9.855\times 10^{-01}$ & $9.683\times 10^{-01}$ & $9.438\times 10^{-01}$ \\
 $1.000\times 10^{-14}$ & $1.337\times 10^{-06}$ & $9.618\times 10^{-01}$ & $9.166\times 10^{-01}$ & $8.524\times 10^{-01}$ \\
 $1.000\times 10^{-13}$ & $2.879\times 10^{-06}$ & $9.039\times 10^{-01}$ & $7.959\times 10^{-01}$ & $6.512\times 10^{-01}$ \\
 $1.000\times 10^{-12}$ & $6.204\times 10^{-06}$ & $7.971\times 10^{-01}$ & $5.958\times 10^{-01}$ & $3.771\times 10^{-01}$ \\
 $1.000\times 10^{-11}$ & $1.337\times 10^{-05}$ & $6.620\times 10^{-01}$ & $3.897\times 10^{-01}$ & $1.703\times 10^{-01}$ \\
 $1.000\times 10^{-10}$ & $2.879\times 10^{-05}$ & $5.310\times 10^{-01}$ & $2.379\times 10^{-01}$ & $6.782\times 10^{-02}$ \\
 $1.000\times 10^{-09}$ & $6.204\times 10^{-05}$ & $4.201\times 10^{-01}$ & $1.408\times 10^{-01}$ & $2.565\times 10^{-02}$ \\
 $1.000\times 10^{-08}$ & $1.337\times 10^{-04}$ & $3.302\times 10^{-01}$ & $8.245\times 10^{-02}$ & $9.514\times 10^{-03}$ \\
 $1.000\times 10^{-07}$ & $2.879\times 10^{-04}$ & $2.576\times 10^{-01}$ & $4.822\times 10^{-02}$ & $3.528\times 10^{-03}$ \\
 $1.000\times 10^{-06}$ & $6.204\times 10^{-04}$ & $2.001\times 10^{-01}$ & $2.813\times 10^{-02}$ & $1.298\times 10^{-03}$ \\
 $1.000\times 10^{-05}$ & $1.337\times 10^{-03}$ & $1.545\times 10^{-01}$ & $1.637\times 10^{-02}$ & $4.787\times 10^{-04}$ \\
 $1.000\times 10^{-04}$ & $2.879\times 10^{-03}$ & $1.181\times 10^{-01}$ & $9.518\times 10^{-03}$ & $1.773\times 10^{-04}$ \\
 $1.000\times 10^{-03}$ & $6.204\times 10^{-03}$ & $8.923\times 10^{-02}$ & $5.508\times 10^{-03}$ & $6.523\times 10^{-05}$ \\
 $1.000\times 10^{-02}$ & $1.337\times 10^{-02}$ & $6.637\times 10^{-02}$ & $3.167\times 10^{-03}$ & $2.395\times 10^{-05}$ \\
 $1.000\times 10^{-01}$ & $2.879\times 10^{-02}$ & $4.813\times 10^{-02}$ & $1.796\times 10^{-03}$ & $8.815\times 10^{-06}$ \\
 $1.000\times 10^{+00}$ & $6.204\times 10^{-02}$ & $3.366\times 10^{-02}$ & $9.940\times 10^{-04}$ & $3.204\times 10^{-06}$ \\
 $1.000\times 10^{+01}$ & $1.337\times 10^{-01}$ & $2.214\times 10^{-02}$ & $5.256\times 10^{-04}$ & $1.143\times 10^{-06}$ \\
 $1.000\times 10^{+02}$ & $2.879\times 10^{-01}$ & $1.300\times 10^{-02}$ & $2.531\times 10^{-04}$ & $3.844\times 10^{-07}$ \\
 $1.000\times 10^{+03}$ & $6.204\times 10^{-01}$ & $5.748\times 10^{-03}$ & $9.349\times 10^{-05}$ & $1.035\times 10^{-07}$ \\
\noalign{\smallskip}\hline
\end{tabular}
\end{table}
Useful surface coverage levels of dust collectors range from $10^{-4}$ to $10^{-2}$,
i.e. a  $ 1 \,{\rm cm}^2$ 
collection surface will contain approximately 100 to $10^4$ particles of $10\,
\mu{\rm m}$ in size. According to Table~\ref{tab:areacum} these particles
constitute about 50\% of the
covered surface (for the size distribution with $\alpha = -3.7$). The other half of the covered surface consists of more finely
dispersed smaller grains.


The COSIMA instrument is sensitive to dust grains of 10\,\micron\ and has 
23 individual collectors. 
The MIDAS instrument is sensitive to submicron-sized dust grains and has 60 individual collectors.
The exposure of  
these collectors to the cometary dust flux and the collection of 
sufficient numbers of dust particles will at the same time lead to 
significant dust coverage of all spacecraft surfaces facing the comet. 
During one dust collection passage of Rosetta at large 
heliocentric distances (2.5 and 3~AU) an individual dust collector may reach 
only a total dust coverage of approximately $2 \times 10^{-3}$ and $7 \times
10^{-4}$, respectively. 
Close to perihelion, however, the comet-facing side of the spacecraft
will be almost completely covered by dust during a single dust collection
passage. 


\subsection{Radiation Environment}
\label{subsec:mmmodel_rad}
In the following we predict the radiation levels received in the vicinity of the comet nucleus due to the presence of dust. It is beyond the
scope of the present work to introduce a detailed wavelength-dependent
model. Instead, values will only be given for the total amount of scattered
visible light and thermally emitted radiation. The crucial parameters for these estimates are the albedo, phase function and
temperature of the dust particles.
We adopt the phase function derived for comet 1P/Halley \cite{divine1981} as
shown in Figure~\ref{fig:phase_fns}, a dust geometric albedo of
4\% \cite{hanner-tedesco1985a}, and a temperature at perihelion of 285~K (see Figure~\ref{fig:dust_temp}).

A priori it is not clear that it is valid to assume that the temperature of a
particle of a given size is independent of its position in the coma,
because the gas molecules impacting on the particle surface give rise to a
heat exchange between the gas and the dust phase. This effect was taken into
account in early models \cite{probstein1968, kitamura1986a},
but was found to be of minor importance later \cite{divine-fechtig1986,
  knollenbergPhD1994}. Neglecting the heat exchange, the temperature of a
dust particle is given by the equilibrium between incident
radiation and thermal emission of the particle. If the coma is optically thin, 
which is shown to be true for the dust coma of 67P/C--G below, the
incident radiation is dominated by solar illumination. Hence it is justified to
treat the particle temperature as independent of the position in the
coma. Assuming furthermore that the physical properties of the particles on
their way from the inner to the outer coma do not change significantly, the
infrared spectrum of dust near the nucleus can be approximated by the
spectrum observed from Earth.  

The thermal emission spectrum of cometary dust particles can only
approximately be represented by the spectrum of a blackbody,
because particles emit efficiently only at wavelengths smaller than
their size. Consequently, the smallest (or very porous) particles
have temperatures much higher than a blackbody at the given heliocentric
distance, and it
is not physically consistent to set the dust temperature to a
constant value independent of particle size. However, for the purpose of
estimating the amount of radiation received by an observer, only the properties
of the dust particles as an ensemble are of importance. 
The relative contribution to the total cross section of particles of a given
size in most of the coma volume is the same as in the ensemble seen by an
Earth-based observer. This was
found with the present model which consistently models the dynamics of
particles of different sizes. To calculate the
intensity received inside the coma it is therefore valid to assign every particle
independently of its size the ensemble properties as measured in Earth-based
observations, which is henceforth done for the optical and thermal
characteristics of the particles. 

Compared with the results given in the previous subsections -- which were strongly
controlled by model parameters only very indirectly accessible to
ground-based observations (e.g. particle size distribution, particle density)
-- the results presented here
only depend on parameters readily measured from ground. For
example the estimates of the scattered visual radiation are governed solely by 
the \afr\ value and the dust phase function. It may, at first glance, be
surprising that the results are not influenced by the 
dust albedo. However, because in the present model the overall dust activity
is adjusted in order to match an observed \afr\ value, the predictions of the
scattered visual radiation inside the coma are merely a scaling of the
Earth-based observations to a different geometry. To give
an idea of the accuracy of the results obtained for the scattered radiation:
The 
present model applied to comet 1P/Halley during the fly-by of Giotto reproduces
the measurements of the Halley Optical Probe experiment
\cite{levasseur-bertaux1986, fulle-levasseur2000} by a factor of only
1.5.  
By contrast, the predictions of the optical thickness and of the thermal radiation by the
dust depend on more parameters, introducing uncertainty: The estimated optical thickness is influenced by the dust
albedo and by the -- poorly constrained -- extinction efficiency, while
the thermal radiation relies on the dust temperature.

Using an extinction efficiency of $q_{\rm ext}=1$ we find for the optical
thickness at perihelion at the subsolar point values of $\tau=0.013$ in the
insolation driven model and $\tau=0.09$ in the jet-model. In the insolation driven model the optical
thickness of the coma along the line of sight from a point at the surface to
the Sun is fairly constant over the
  comet surface and the results can well be approximated by the simplified model in \cite{mueller-green2002a}. In the jet model
the optical thickness decreases towards the terminator by a factor of $5$.

Since the extinction efficiency $q_{\rm ext}$ is little constrained and can assume values in the range
$q_{\rm ext} \approx 1-2$ we estimate that at perihelion the optical thickness
at the comet surface is typically in the range $\tau \approx
0.01-0.03$. Within jets it can reach values up to $\tau \approx 0.2$. Since the optical thickness is proportional to
\afr, the results can easily be scaled to scenarios with different levels of
cometary activity.

\begin{figure}[h]
\includegraphics[width=0.5\textwidth]{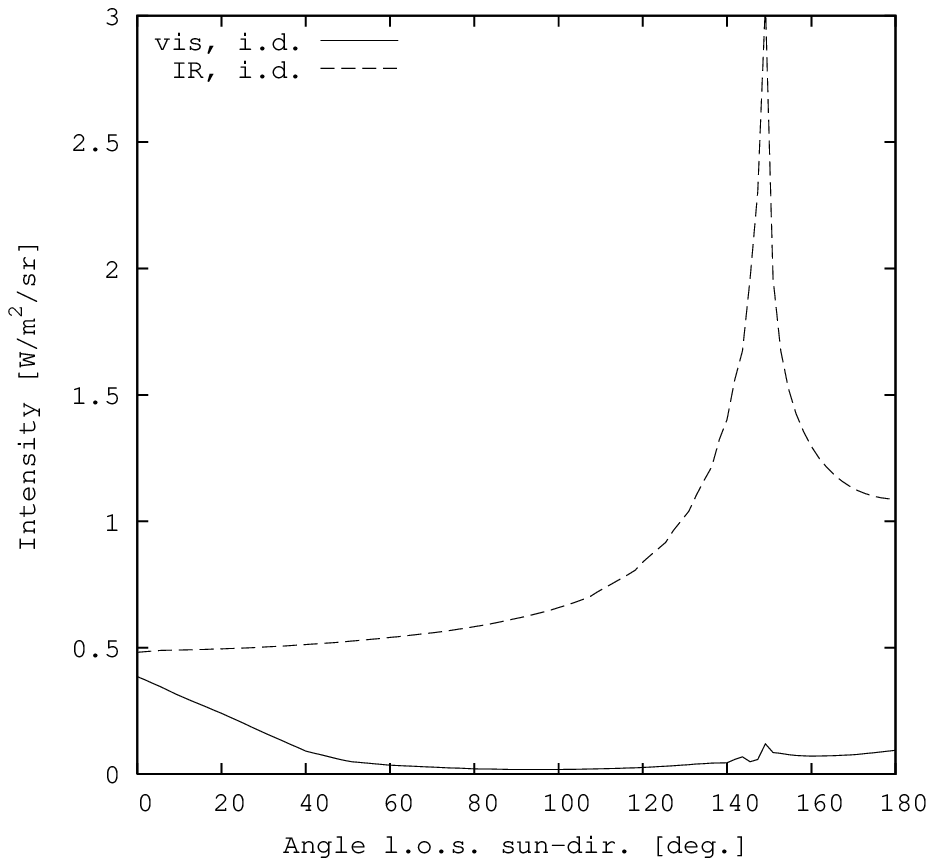}
\includegraphics[width=0.5\textwidth]{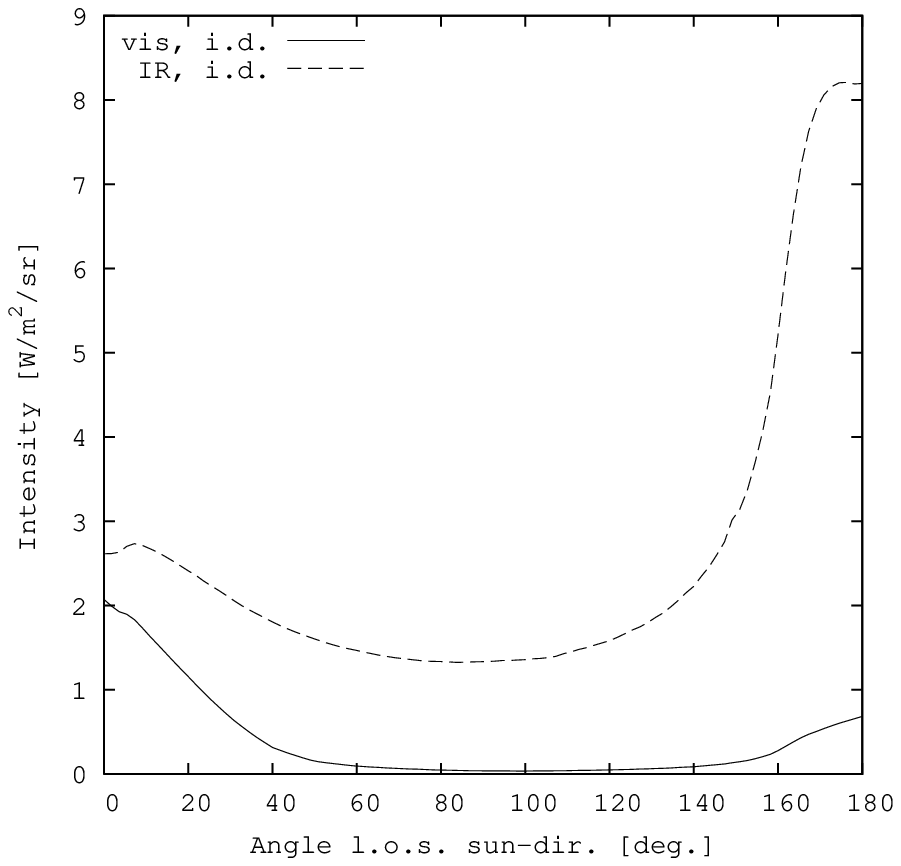}
\caption{Scattered sunlight and thermal infrared radiation at perihelion received from a line of sight by an observer above the
  subsolar point at 4 km from the nucleus centre. Only the contribution by the
  dust is considered, in contrast to radiation from the nucleus or direct
  sunlight. The nominal size distribution $\alpha = -3.7$ was used. Left:
  insolation driven model,  right: jet model.}
\label{fig:rad_los}
\end{figure}

The intensity received by an observer at 4 km (2 comet radii) from the nucleus centre over the subsolar point for different lines of sight is shown in Figure~\ref{fig:rad_los}.
For the insolation driven model, the observed visible intensity reaches maxima
for lines of sight directed towards the Sun and the nucleus (phase angle
$180^\circ$ and $0^\circ$, respectively). These peaks are not
found in the infrared intensity received in the insolation driven model, which
shows that they are due to the forward and backward scattering peaks of the
dust phase function (Figure~\ref{fig:phase_fns}). In the insolation driven
model the intensities, both for visible and infrared radiation, reach maxima
for the lines of sight that touch the nucleus tangentially (at about
$150^\circ$). In the jet-model these peaks are not present because most of the
dust is concentrated to a narrow region over the subsolar point. This also
explains the maxima of the infrared intensity in the jet-model for the lines
of sight in the Sun- and the nucleus-direction that are not observed in the
insolation driven model. Correspondingly, the maxima for the visible radiation
on these lines of sight are more pronounced in the jet-model than in the
insolation driven model. 

\begin{figure}[h]
\includegraphics[width=.5\textwidth]{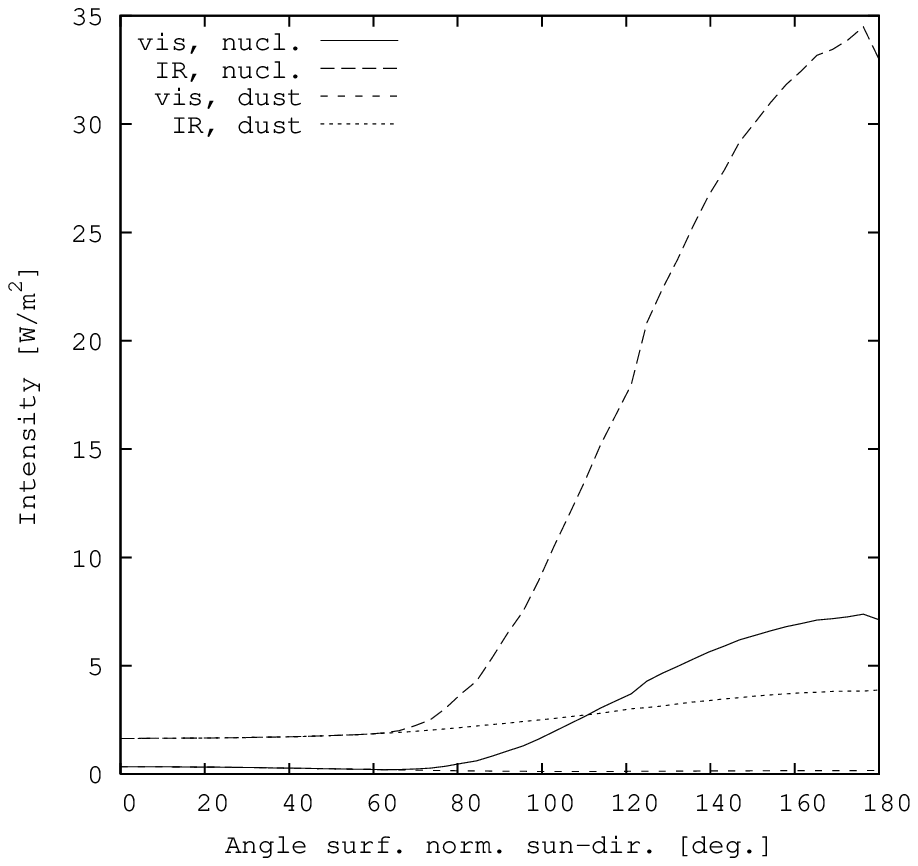}
\includegraphics[width=.5\textwidth]{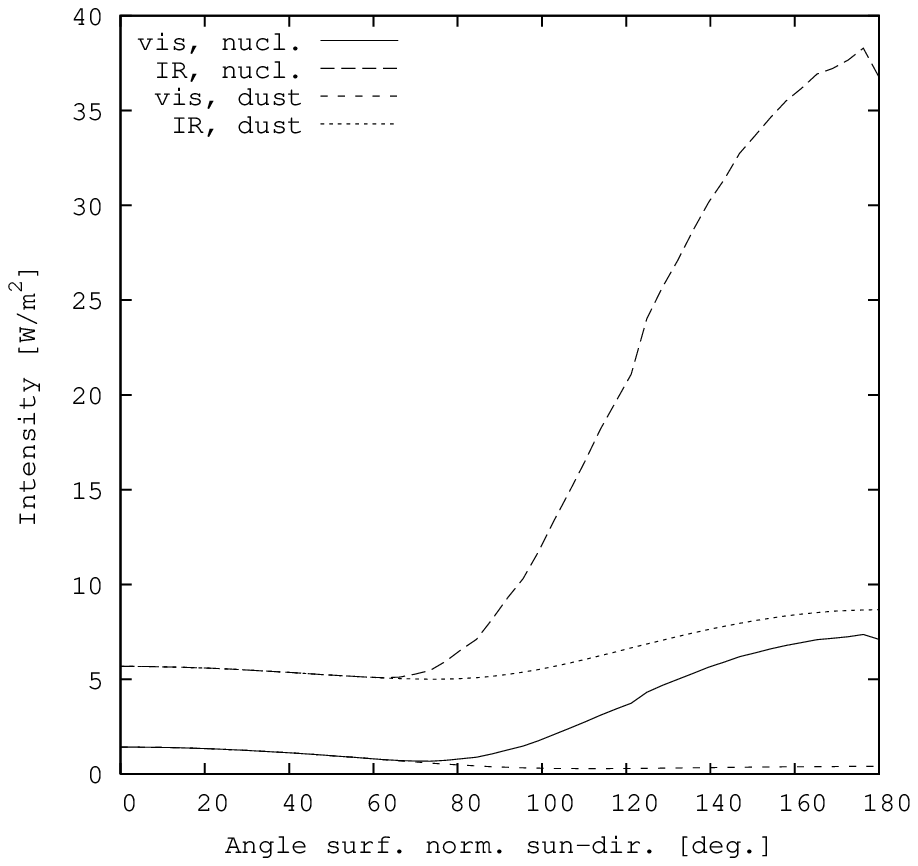}
\caption{Radiation at perihelion from all directions on a surface above the
  subsolar point at 4 km from the nucleus centre. Both curves with and without
  the contribution from the nucleus are shown, but no direct sunlight is considered. The nominal size distribution $\alpha = -3.7$ was used. Left:
  insolation driven model, right: jet model.}
\label{fig:rad_plane}
\end{figure}

The total intensity received by a surface depends on its orientation. To give
an example, we consider a plane surface at
4 km over the subsolar point (see
Figure~\ref{fig:rad_plane}) taking into account also the radiation from the
nucleus. The intensities emitted or scattered by the nucleus dominate over
those from the dust; and, since the nucleus is a dark object, the infrared
radiation dominates the total intensity. Note that the nucleus temperature in
the model is calculated as that of a pure icy surface. Thus the given results for
the infrared radiation received from the nucleus are lower limits. If the
major part of the comet surface is inactive, the surface temperature and
consequently the infrared flux will be much higher.

\section{Summary and Conclusions}
\label{sec:summary}
We have summarised the present knowledge on the dust emitted by Rosetta target
comet 67P/Churyumov-Gerasimenko and discussed methods and results of modelling
it. The observational data can be described as follows: The dust and gas production rates peaked about 30
days after perihelion passage during three past apparitions. The
coma was characterised by azimuthal brightness variations probably resulting
from active areas on the surface. Close to the comet orbit, a distinct
line-shaped feature prevailed at least from shortly after perihelion 2002
until 2006, beyond aphelion. It is interpreted as a pronounced antitail due to
the low inclination of the comet orbital plane towards the ecliptic.
The colour temperatures in both the coma and the trail were higher than the
blackbody equilibrium temperature at the concerned heliocentric distance.

We have reviewed different approaches to derive parameters of the cometary
dust production by modelling images of the tail and/or trail, and the results
obtained by their application to observations of \cg. We found that due to the
large diversity of the derived parameter values it is at present not possible
to formulate a consistent picture of the \cg\ dust activity and its time
evolution. We identified the need to find a common model that satisfies all
available observations.

Using the ESA Cometary
Dust Environment Model we have predicted the terminal speeds of dust,
the dust flux in the coma and along sample trajectories of the spacecraft, and
the radiation flux in the coma. We listed results for three mass
distributions and five heliocentric distances. We also considered different
surface activity distributions, among which one implying a homogeneous surface of the
nucleus and one assuming all dust to be emitted by a single active area. The
former yielded estimates of quantities averaged over space and time, while the
latter served to derive upper limits for local quantities.

\begin{acknowledgement}
M. M. gratefully acknowledges funding by EDS Operations Services GmbH. 
We wish to thank R.~Schulz et al. \cite{schulz-stuewe2004a,
  schulz-stuewe2004b} for making their data available to us. 
\end{acknowledgement}

\bibliographystyle{spphys}
\bibliography{/usr4/users/jagarwal/Latex/refs}

\end{document}